\documentclass[review]{elsarticle}
\usepackage{amsmath,amssymb}
\usepackage{lipsum,soul}
\usepackage{bm}
\usepackage{graphicx}
\usepackage{graphics}
\usepackage{amsmath}
\usepackage{array}
\usepackage[caption=false]{subfig}
\usepackage{xcolor}
\usepackage{subfig}
\usepackage{hyperref}
\usepackage[capitalise]{cleveref}
\usepackage{textcomp} 

\usepackage{epstopdf}
\usepackage{cases}
\usepackage{amssymb}
\usepackage{multirow}
\usepackage{siunitx}
\usepackage{cases}
\usepackage{tabularx}
\usepackage{nomencl}
\usepackage[top=1.1in, bottom=1.1in, left=1.0in, right=1.0in]{geometry}
\usepackage{tikz}
\usepackage{comment}
\usepackage{lineno,hyperref}
\newtheorem{remark}{Remark}
\modulolinenumbers[5]
\usepackage{booktabs}

\begin{document}
\title{Local reactive boundary scheme for lattice Boltzmann method}
\author[add1]{Long Jv}
\ead{aiad520@hust.edu.cn}
\author[add1]{Chunhua Zhang}
\ead{zchua@hust.edu.cn}
\author[add1]{Zhaoli Guo\corref{cor1}}
\ead{zlguo@hust.edu.cn}
\cortext[cor1]{Corresponding author}
  \address[add1]{State Key Laboratory of Coal Combustion, Huazhong University of Science and Technology,Wuhan, 430074, China}

\date{\today}
\begin{abstract}

In this paper, a boundary scheme is proposed for the two-dimensional five-velocity (D2Q5) lattice Boltzmann method with heterogeneous surface reaction, in which the unknown distribution function is determined locally based on the kinetic flux of the incident particles. Compared with previous boundary schemes, the proposed scheme has a clear physical picture that reflects the consumption and production in the reaction. Furthermore, the scheme only involves local information of boundary nodes such that it can be easily applied to complex geometric structures. In order to validate the accuracy of the scheme, some benchmark tests, including the convection-diffusion problems in straight and inclined channels are conducted. Numerical results are in excellent agreement with the analytical solutions, and the convergence tests demonstrate that second-order spatial accuracy is achieved for straight walls, and the order of accuracy is between 1.5 and 2.0 for general inclined walls. Finally, we simulated the density driving flow with dissolution reactions in a two-dimensional cylindrical array, and the results agree well with those in previous studies.

\end{abstract}
\maketitle
\section{Introduction}

Fluid flows with fluid-solid surface reactions are prevalent phenomenas in many natural and industrial applications, such as removal of subsurface
contaminant~\cite{garcia2010contaminant}, ion transport in fuel cells~\cite{willingham2008evaluation,jaouen2011recent} and well bore acidization~\cite{nasr2007chemical,cohen2008pore}. Mathematically, such processes can be described by the convection-diffusion equation (CDE) coupled with the Navier-Stokes (NS) equations~\cite{guo2007thermal} with reactive boundary conditions. In the past few years, many numerical simulations have been conducted to understand these processes using the traditional numerical methods~\cite{asinari2007direct, zeiser2001cfd, li2001heterogeneous,yoon2015lattice}. But most of these processes occur in a physical domain with extremely complex geometries, the conventional numerical methods suffered from many difficulties ~\cite{succi2001lattice,aidun2010lattice}.

In the past two decades, the lattice Boltzmann equation (LBE) method that originates from kinetic theory has become an effective tool for simulating various fluid flow problems~\cite{guo2002zl, chen1998lattice, sheikholeslami2016lattice}. Compared with the traditional methods, LBE has a clear physical interpretation and can easily deal with complex boundaries~\cite{guo2008lattice}. Some LBE methods have been developed for the convection diffusion problems with surface reaction~\cite{verhaeghe2006lattice, ginzburg2005generic,zhang2002boundary}. Particularly, the linear heterogeneous surface reactions described by the general Robin boundary condition have received much attention~\cite{he2000lattice,kang2006lattice,kamali2012mr,li2013multiple,huber2014new,kang2006lattice,zhang2012general,chen2013improved,Huang2015boundary,Meng2016boundary}.

Among these LBE models, the boundary schemes are usually limited to straight boundaries~\cite{he2000lattice}, or stationary boundaries~\cite{kang2006lattice,kamali2012mr,li2013multiple}, and the spatial accuracy of most schemes mentioned above are not clear ~\cite{huber2014new,kang2006lattice}.
For instance, Zhang \emph{et al}~\cite{zhang2012general} developed a general bounce-back scheme to solve the general Robin boundary condition, in which the normal direction on the boundary is approximated by the lattice link direction. And the normal derivative in the boundary condition is discretized by a first-order finite-difference scheme, such that the scheme is only of first-order accuracy and the computation is non-local.
Later, Chen \emph{et al}~\cite{chen2013improved} improved the scheme in which the finite-difference approximation is performed along the boundary normal direction. But the accuracy remains first-order and the computation is still non-local.
Recently, Huang \emph{et al}~\cite{Huang2015boundary} employed an asymptotic analysis technique and constructed two boundary schemes for the general Robin boundary condition on straight and curved boundaries, respectively. The schemes is of second-order accuracy for straight boundaries while first order accuracy for curved boundaries. Although their schemes involve only the current lattice node, the linear combination of the local post-collision distribution functions is a bit complicated and numerical instability may occur for curved boundaries when the relaxation time is close to 1.0.
Most recently, Meng \emph{et al}~\cite{Meng2016boundary} developed a single-node boundary scheme which computes the scalar gradient in the boundary condition by the moment of the non-equilibrium distribution functions. However, this scheme needs to distinguish whether the boundary is straight or curved in advance, which is not suitable for complex geometries.
In brief, the unknown distributions in the above schemes are all obtained by solving the macroscopic boundary conditions, so the spatial accuracy may be insufficient because of the first-order finite-difference for the normal derivative ~\cite{zhang2012general,chen2013improved}, or the computational procedures are very complicated and the physical process is not clear ~\cite{Huang2015boundary,Meng2016boundary}. In addition, most of these schemes are designed for straight and the curved boundaries separately~\cite{chen2013improved,Huang2015boundary,Meng2016boundary}. Therefore, accurate uniform boundary schemes for both straight and curved boundaries with clear physics are still desired.

 For this purpose, a localized boundary scheme for a two-dimensional LBE model with linear heterogeneous reactions is proposed, which takes the same formulation for both straight and curved walls. In the scheme, the unknown distribution functions are constructed with clear physics, which reflects the loss and gain at the boundary due to the reaction. The scheme can also be easily extended to three-dimensional problems.

The rest of this article is organized as follows. In Sec.\ref{sec:2}, the two-dimensional LBE model for the CDE with linear heterogeneous surface reactions is presented; In Sec.\ref{sec:3}. several numerical simulations are conducted to evaluate the performance of the present scheme, and finally a conclusion is drawn in Sec.4.

\section{LBE models and boundary conditions}\label{sec:2}

\subsection{LBE models for convection-diffusion equation}

In the present study, we focus on the transport processes in two dimensions (2D) for simplicity. The two dimensions (2D) CDE can be expressed as follows~\cite{li2017lattice}:
\begin{equation}
\frac{\partial C}{\partial t} + \nabla \cdot \left(\bm{u}{C}\right)= D \nabla^2 C ,
\label{eq:1}
\end{equation}
where $\bm{u}=(u,v)$ is the velocity vector and $u$,$v$ are the components of the velocity in the x-direction and y-direction respectively. $C$ is the conserved scalar variable such as concentration, $D$ represents the diffusion coefficient. The evolution equation of the LBE model for Eq.~\eqref{eq:1} can be written as~\cite{zhang2012general}
\begin{equation}
g_i \left(\bm{x}+\bm{c_i} \delta_t,t+ \delta_t\right)-g_i \left(\bm{x},t\right)=-\frac{1}{\tau_s}\left[g_i \left(\bm{x},t\right)-g_i^{\left(eq\right)} \left(\bm{x},t\right)\right],
\label{eq:2}
\end{equation}
where $g_i$ is the particle distribution function at position $x$ and time $t$, $\delta_t$ is the time step. $\tau_s$ is the dimensionless relaxation time and $g_i^{\left(eq\right)}$ represents the equilibrium distribution functions. Here we employ the two-dimensional-five-velocity (D2Q5)~\cite{qian1992lattice} model for presentation. The equilibrium distribution function is~\cite{kang2007improved, Huang2016second}
\begin{equation}
g_i^{\left(eq\right)}=\omega_iC\left[1+\frac{\bm{c_i \cdot u}}{c_s^2}\right],
\label{eq:3}
\end{equation}
where $\omega_i$ is the weight coefficient defined as $\omega_0=1/3$, $\omega_{1-4}=1/6$ in D2Q5 model. $c_s$ is the lattice sound speed defined as $c_s=c/\sqrt{3}$ and $\bm{c_i}$ represent the discrete velocity, which is defined as:
 \begin{equation}
c_i=
\begin{cases}
c(0,0),& i=0,\\
c(cos[(i-2)\pi/2],sin[(i-2)\pi/2]),& i=1-4,
\end{cases}
\label{eq:4}
\end{equation}
where $c=\delta_x/\delta_t$ is the particle speed with $\delta_x$ being the lattice spacing.

To recover Eq~\eqref{eq:1} by the Chapman-Enskog expansion, the diffusion coefficient is defined as (see the appendix for more details) :
\begin{equation}
D=c_s^2(\tau_s-0.5)\delta_t.
\label{eq:5}
\end{equation}
The scalar variable can be calculated by
 \begin{equation}
C=\sum_i g_i.
\label{eq:6}
\end{equation}

\subsection{Boundary scheme for the linear heterogeneous reactions}

The linear heterogeneous reactions at the fluid-solid interface can be described as~\cite{mostaghimi2016numerical,liu2017pore,Meng2016boundary}
 \begin{equation}
D\frac{\partial C_b}{\partial \bm{n}}=k_r(C_b-C_{eq}),
\label{eq:7}
\end{equation}
where $D$ represents the diffusion coefficient. $\bm{n}$ is the unit normal vector on the boundary pointing into the computational domain. $k_r$ is the rate of the chemical reactions. $C_b$ is the concentration at the boundary and $C_{eq}$ is the equilibrium concentration of the reaction. The left-hand side of Eq.~\eqref{eq:7} represents the flux due to diffusion and the right-hand represents the surface reaction.

The modified bounce-back scheme~\cite{guo2008lattice,gallivan1997evaluation,zou1997pressure,he1997analytic} can be expressed as
\begin{equation}
g_i=g_{i^{'}} ,
\label{eq:8}
\end{equation}
 where $i^{'}$ represents the opposite direction of $i$.
 It means that all the fluid particles bounce back when they come to the wall, leading to mass conserved strictly. But when the reactions occur at the boundary, part of the fluid particles are consumed, it may not mass conserved any more. Thus, we can introduce a parameter $\alpha_i$, and the unknown distribution functions on the boundary can be expressed  as
 \begin{equation}
g_i=\alpha_i g_{i^{'}},
\label{eq:9}
\end{equation}
which represents that after the reaction, only $\alpha_i g_{i^{'}}$ left. When the reaction is reversible, Eq.~\eqref{eq:9} can be amended  as
 \begin{equation}
g_i=\alpha_i g_{i^{'}}+\beta_i,
\label{eq:10}
\end{equation}
where $\beta_i$ represents the part of reverse reaction. 
In what follows, we will identify the parameter $\alpha_i$ and $\beta_i$.

\begin{figure}
 \centering
 \includegraphics[width=0.45\textwidth]{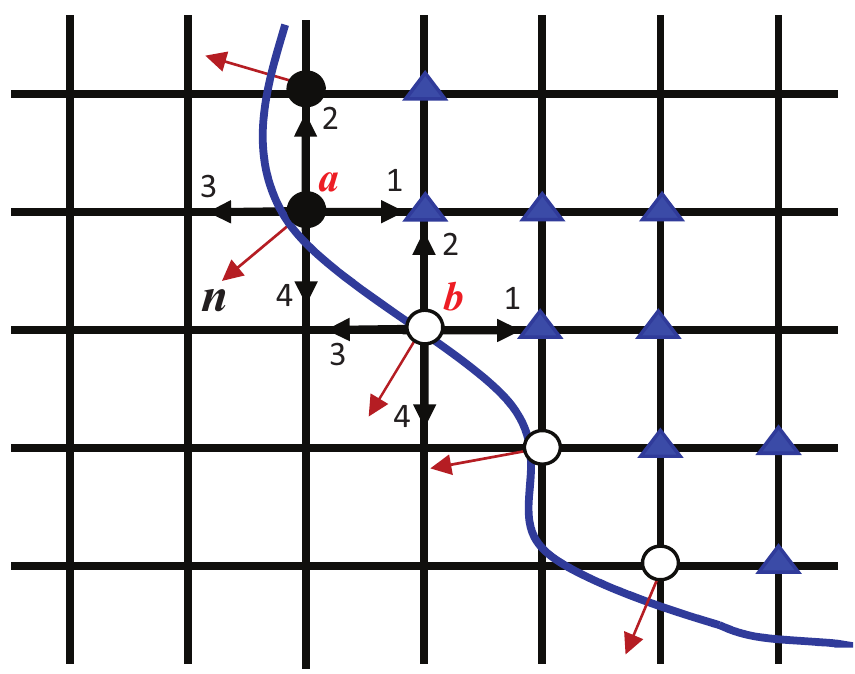}~
 \caption{Schematic illustration for the curved boundary. Blue lines: physical boundary}
 \label{curvedtwall}
\end{figure}

In this study, we use the zig-zag boundaries to approximate the physical boundaries~\cite{liu2015pore, lei2017pore,kang2010lattice}. The collision and streaming occur at all the boundary nodes which are similar to the modified bounce-back. As shown in \cref{curvedtwall}, the dots represent the computational boundary and the triangles represent the internal of solid. For the D2Q5 model, the lattice nodes which are used to approximate the physical boundary can be divide into two types. A sort of them only have one distribution function remains to be determined, which are marked with the black dots. And others have two distribution functions remain to be determined by the boundary condition which are marked with the hollow dots.

For the points on the boundary, the total flux along the normal direction in the discrete case can be expressed as~\cite{tao2016investigation}:
 \begin{equation}
  \begin{split}
& \bm{j_n}=\bm{j}_{\bm{n}D}+\bm{j}_{\bm{nw}}=\sum_{i}\bm{n\cdot c_i}g_i,
\end{split}
\label{eq:11}
\end{equation}
where $\bm{j}_{\bm{n}D}$ represents the flux due to diffusion, and $\bm{j}_{\bm{nw}}$ represents the flux due to convection, which can be expressed as~\cite{Meng2016boundary}
\begin{equation}
  \begin{split}
&  \bm{j}_{\bm{nw}}=\bm{n}\cdot\bm{w}C_b,
\end{split}
\label{eq:12}
\end{equation}
where $\bm{w}$ represents the velocity of wall movement, and $C_b$ represents the concentration at the wall.

On the right hand of Eq.~\eqref{eq:11}, some distribution functions which in the opposite direction are already known, such as $g_2$ and $g_4$ at the point $'a'$ in \cref{curvedtwall}. Substituting Eq. (12) into Eq. (11) leads to:
\begin{equation}
  \begin{split}
      \bm{j}_{\bm{n}D}=\sum_{k}\bm{n\cdot c_k}g_k-\epsilon-\bm{n}\cdot\bm{w}C_b,
  \end{split}
\label{eq:13}
\end{equation}
where $k$ represents the direction where the particles will react with the wall, and $\epsilon$ represents the flux due to the distribution functions which are already known. If the solid lattice nodes that are marked by triangles are denoted by s(x)=1 while the others are denoted by s(x)=0, $\epsilon$ can be expressed as follow:
\begin{equation}
  \begin{split}
      &\epsilon=\sum_{i=1\sim4}\left[\bm{n\cdot c_i}g_i-\bm{n\cdot c_i}s(\bm{x}+\bm{c_i}\delta_t)(g_i-g_{i^{'}})\right],
  \end{split}
\label{eq:14}
\end{equation}
$D{\partial C_b}/{\partial \bm{n}}$ in Eq.~\eqref{eq:7} represents the diffusion flux in the continuous case and $\bm{j}_{\bm{n}D}$ represents the diffusion flux in the discrete case, then we can obtain:
\begin{equation}
  \begin{split}
      \bm{j}_{\bm{n}D}=-\gamma D\frac{\partial C}{\partial \bm{n}}
  \end{split}
\label{eq:15}
\end{equation}
where $\gamma$ represents the parameter introduced by the discrete process. Through the Chapman-Enskog expansion we can obtain (see the appendix for more details):
\begin{equation}
  \begin{split}
      \gamma=\frac{\tau}{\tau-0.5},
  \end{split}
\label{eq:16}
\end{equation}
Substituting Eq.~\eqref{eq:13} and Eq.~\eqref{eq:7} into Eq.~\eqref{eq:15}, we can obtain:
\begin{equation}
  \begin{split}
      \sum_{k}\bm{n\cdot c_k}g_k=-\gamma k_r(C_b-C_{eq})-\epsilon+(\bm{n}\cdot\bm{w})C_b
  \end{split}
\label{eq:17}
\end{equation}

Next we take $g_3$ at the point $'b'$ in \cref{curvedtwall} for example to derive $\alpha_3$, and $\beta_3$. Then Eq.~\eqref{eq:15} can be expressed as
\begin{equation}
  \begin{split}
      (\bm{n\cdot c_3})(\alpha_3g_1+\beta_3-g_1)+(\bm{n\cdot c_4})(g_4-g_2)=-\gamma k_r(C_b-C_{eq})-\epsilon+(\bm{n}\cdot\bm{w})C_b,
  \end{split}
\label{eq:18}
\end{equation}
In Eq.~\eqref{eq:18}, both $\alpha_3$, $\beta_3$, $g_4$ and $C_b$ are unknown. Thus some other equations are supplemented~\cite{,zhang2012general}:
\begin{align}\label{eq:19}
    &g_3=2\omega_3C_b-g_{1},\\
    &g_4=2\omega_4C_b-g_2,\\
    &\beta_3(C_{eq}=0)=0,
\end{align}
where Eq.(21) means that if the reaction is not reversible ($C_{eq}=0$), $\beta_3=0$. Substituting Eqs.(19) and (20) to Eq.~\eqref{eq:18} we can obtain:
\begin{equation}
  \begin{split}
      2\omega_3(\bm{n\cdot c_3})(\alpha_3-1)g_1-2\omega_3(\bm{n\cdot c_2})(\alpha_3+1)g_1-2\omega_3(\bm{n\cdot c_2})\beta_3+2\omega_3(\bm{n\cdot c_3})\beta_3\\
      =(\bm{n}\cdot\bm{w}-\gamma k_r)[(\alpha_3+1)g_1+\beta_3]+2\omega_3\gamma k_rC_{eq}-4\omega_3(\bm{n\cdot c_2})g_2-2\omega_3\epsilon.
  \end{split}
\label{eq:22}
\end{equation}
Substituting Eq.(21) to Eq.~\eqref{eq:22} we can obtain
\begin{equation}
  \begin{split}
      &2\omega_3(\bm{n\cdot c_3})(\alpha_3-1)g_1-2\omega_3(\bm{n\cdot c_2})(\alpha_3+1)g_1=(\bm{n}\cdot\bm{w}-\gamma k_r)[(\alpha_3+1)g_1+\beta_3]\\
      &-4\omega_3(\bm{n\cdot c_2})g_2-2\omega_3\epsilon.
  \end{split}
\label{eq:23}
\end{equation}
The rest of Eq.~\eqref{eq:22} can be expressed as:
\begin{equation}
  \begin{split}
      &2\omega_3(\bm{n\cdot c_2})\beta_3+2\omega_3(\bm{n\cdot c_3})\beta_3=(\bm{n}\cdot\bm{w}-\gamma k_r)\beta_3+2\omega_3\gamma k_rC_{eq}.\\
  \end{split}
\label{eq:24}
\end{equation}
From Eqs.~\eqref{eq:23} and \eqref{eq:24} we can derive that:
\begin{equation}
  \begin{split}
      \alpha_3=\frac{\zeta+2\omega_3(\bm{n\cdot c_3}+\bm{n\cdot c_2}) -4\omega_3\bm{n\cdot c_2}g_2/g_1-2\omega_3\epsilon/g_1}{-\zeta+2\omega_3(\bm{n\cdot c_3}-\bm{n\cdot c_2})},\qquad
  \end{split}
\label{eq:25}
\end{equation}
\begin{equation}
  \begin{split}
      \beta_3=\frac{2\omega_3\gamma k_rC_{eq}}{-\zeta+2\omega_3(\bm{n\cdot c_3}-\bm{n\cdot c_2})},
  \end{split}
\label{eq:26}
\end{equation}
where $\zeta=\bm{n}\cdot\bm{w}-\gamma k_r$. $\alpha_4$ and $\beta_4$ for $g_4$ at the point $'b'$ can also be obtained in the similar process. We put the last two terms into $\beta_i$ for brief and numerical stability,
$\alpha_i$ and $\beta_i$ can be expressed in a more general form:
\begin{equation}
  \begin{split}
      & \alpha_i=\frac{\zeta+2\omega_i(\bm{n\cdot c_i}+\sigma)}{-\zeta+2\omega_i(\bm{n\cdot c_i}-\sigma)},\quad
      \beta_i=\frac{2\omega_i\gamma k_rC_{eq}-4\omega_i\lambda-2\omega_i\epsilon}{-\zeta+2\omega_i(\bm{n\cdot c_i}-\sigma)},
  \end{split}
\label{eq:27}
\end{equation}
with
 \begin{equation}
  \begin{split}
     & \sigma=\sum_{k=1-4}\left[s(\bm{x}+\bm{c}_k\delta t)(\bm{n\cdot c}_k)\right]+(\bm{n\cdot c}_{i}),\\
     &\lambda=\sum_{k=1-4}\left[s(\bm{x}+\bm{c}_k\delta t)(\bm{n\cdot c}_k)g_k\right]+(\bm{n\cdot c}_{i})g_i,\\
     &\epsilon=\sum_{k=1-4}\left[(\bm{n\cdot c}_k)g_k-(\bm{n\cdot c}_k)s(\bm{x}+\bm{c}_k\delta t)(g_k-g_{k^{'}})\right],
  \end{split}
\label{eq:28}
\end{equation}
In particular, for a straight and stationary wall, Eq.~\eqref{eq:27} can be written as
\begin{equation}
  \begin{split}
      & \alpha_i=\frac{-\gamma k_r+2\omega_i(\bm{n\cdot c_i})}{\gamma k_r+2\omega_i(\bm{n\cdot c_i})},\qquad
      \beta_i=\frac{2\omega_i\gamma k_rC_{eq}}{\gamma k_r+2\omega_i(\bm{n\cdot c_i})},
  \end{split}
\label{eq:29}
\end{equation}


\begin{remark}
If there is no reaction occurs ($k_r=0$) on the boundaries, Eq.~\eqref{eq:29} can be expressed as:
\begin{equation}
  \begin{split}
      & \alpha_i=1,\qquad
      \beta_i=0,
  \end{split}
\label{eq:30}
\end{equation}
in this time, our scheme turn to the modified bounce-back scheme.
\end{remark}

\begin{remark}
  When the rate of reaction is infinite ($k_r=\infty$), which means the reactions are always in equilibrium ($C_{eq}$), Eq.~\eqref{eq:29} can be expressed as:
\begin{equation}
  \begin{split}
      & \alpha_i=-1,\qquad
      \beta_i=2\omega_i C_{eq},
  \end{split}
\label{eq:31}
\end{equation}
then our scheme turn into the scheme developed for Dirichlet boundary condition.
\end{remark}

\section{NUMERICAL RESULTS AND DISCUSSION}\label{sec:3}

In this section, we will carry out some bunchmarks to demonstrate the accuracy of the proposed boundary scheme, including a one dimensions (1D) convection-diffusion problem, and a convection-diffusion problem in an inclined channel. The spacial accuracy of the present scheme is analyzed. In the end, the density-driven flows with dissolution in porous media are studied using this scheme and compared with previous results.

\subsection{One-dimensional convection-diffusion system}

The one-dimensional (1D) unsteady convection-diffusion system can be explained mathematically as follows:
 \begin{equation}
\partial_tC+u\partial_xC=D\partial_x^2C,
\label{eq:32}
\end{equation}
where the concentration $C$ is a function of the position $x$ and time $t$. $u$ is the velocity in the $x$ direction. This problem can be characterized by the P\'{e}clet number defined as $\text{Pe}=uL/D$. The initial conditions are as follow:
 \begin{equation}
C(x,0)=0, \qquad 0\leq x \leq L
\label{eq:33}
\end{equation}
The boundary conditions are given as
  \begin{equation}
  \begin{split}
     & uC(0,t)-D\partial_x C(0,t)=uA_f, \qquad t>0\\
     &\partial_x C(L,t)=0 \qquad t>0,
  \end{split}
\label{eq:34}
\end{equation}
where $L$ is the length of the computational domain.

For this case, we impose the periodic boundary conditions on the upper and bottom boundaries to extend this problem into two dimensions. The analytical solution of this problem can be written as~\cite{Meng2016boundary}
   \begin{equation}
  \begin{split}
    & C^{'}(x,t)=A_f \{\frac{1}{2}erfc\left(\frac{x-ut}{2\sqrt{Dt}}\right)+\sqrt{\frac{u^2t}{\pi D}}exp\left[-\frac{(x-ut)^2}{4Dt}\right]\\
     &\qquad \qquad -\frac{1}{2}\left(1+\frac{ux}{D}+\frac{u^2t}{D}\right)exp\left(\frac{ux}{D}\right)erfc\left(\frac{x+ut}{2\sqrt{Dt}}\right)\}
  \end{split}
\label{eq:35}
\end{equation}

\begin{figure}
     \centering
     \subfloat[]{\includegraphics[width=0.45\textwidth]{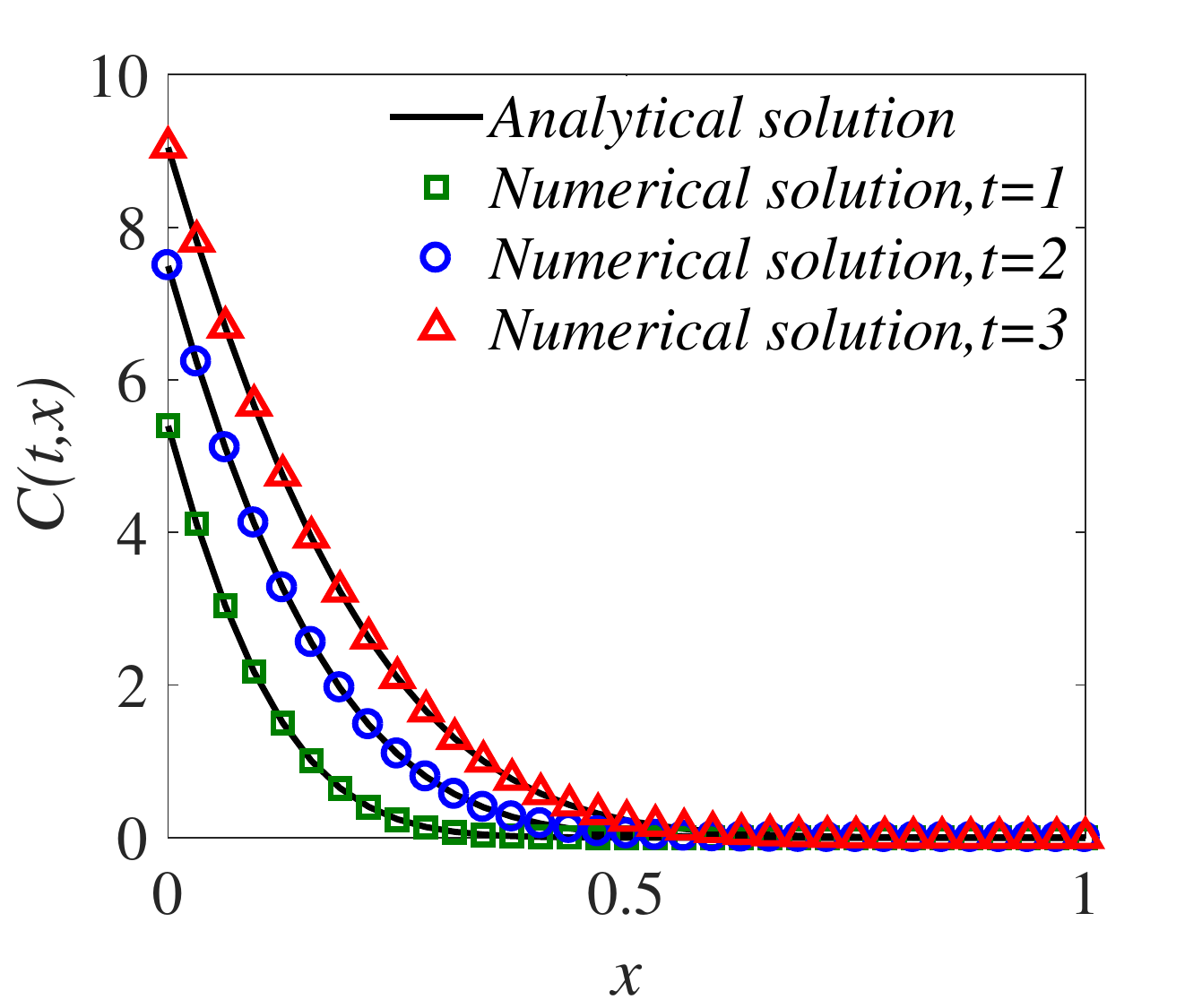}}~~
     \subfloat[]{\includegraphics[width=0.45\textwidth]{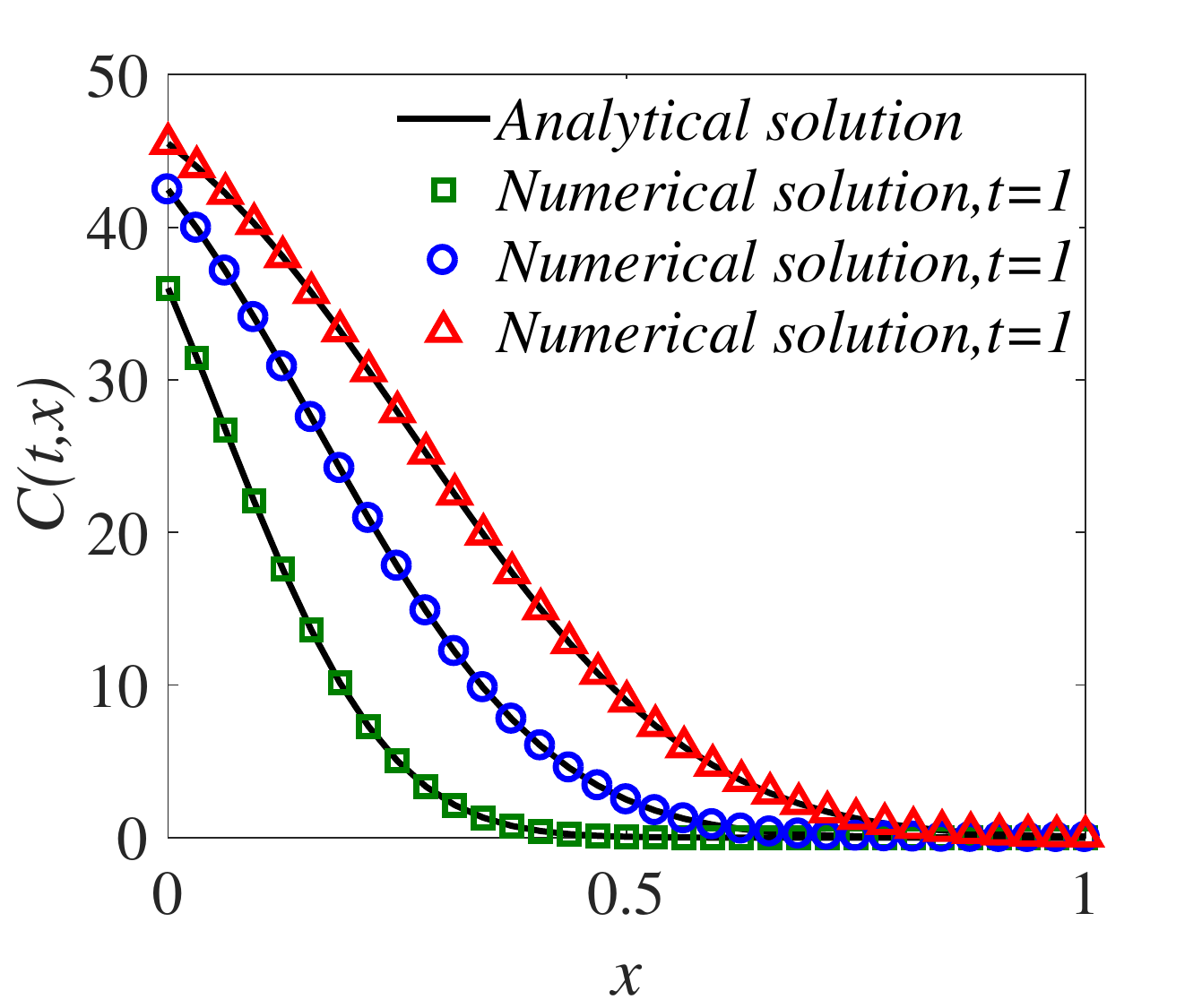}}~~
     \caption{Concentration profiles at different times for different P´eclet numbers. (a) Pe = 1, (b) Pe = 10.}
     \label{case1e1}
\end{figure}

 In our study below, two representative cases are simulated, i.e., $\text{Pe}=1$ and $\text{Pe}=10$. We set $D=0.01, A_f=50$ and $L=1$. The relaxation time $\tau_s$ is set to be 0.53. All the parameters mentioned above are chosen to be same with those used in Ref~\cite{Huang2015boundary}.  he value of the time step is $\delta _t=\delta_x^2(\tau_s-0.5)/3D$ according to Eq.~\eqref{eq:5}. The grid size used in the simulations is 32×32. Good agreement with the analytical solutions are shown in~\cref{case1e1}.

\begin{figure}
     \centering
    {\includegraphics[width=0.49\textwidth]{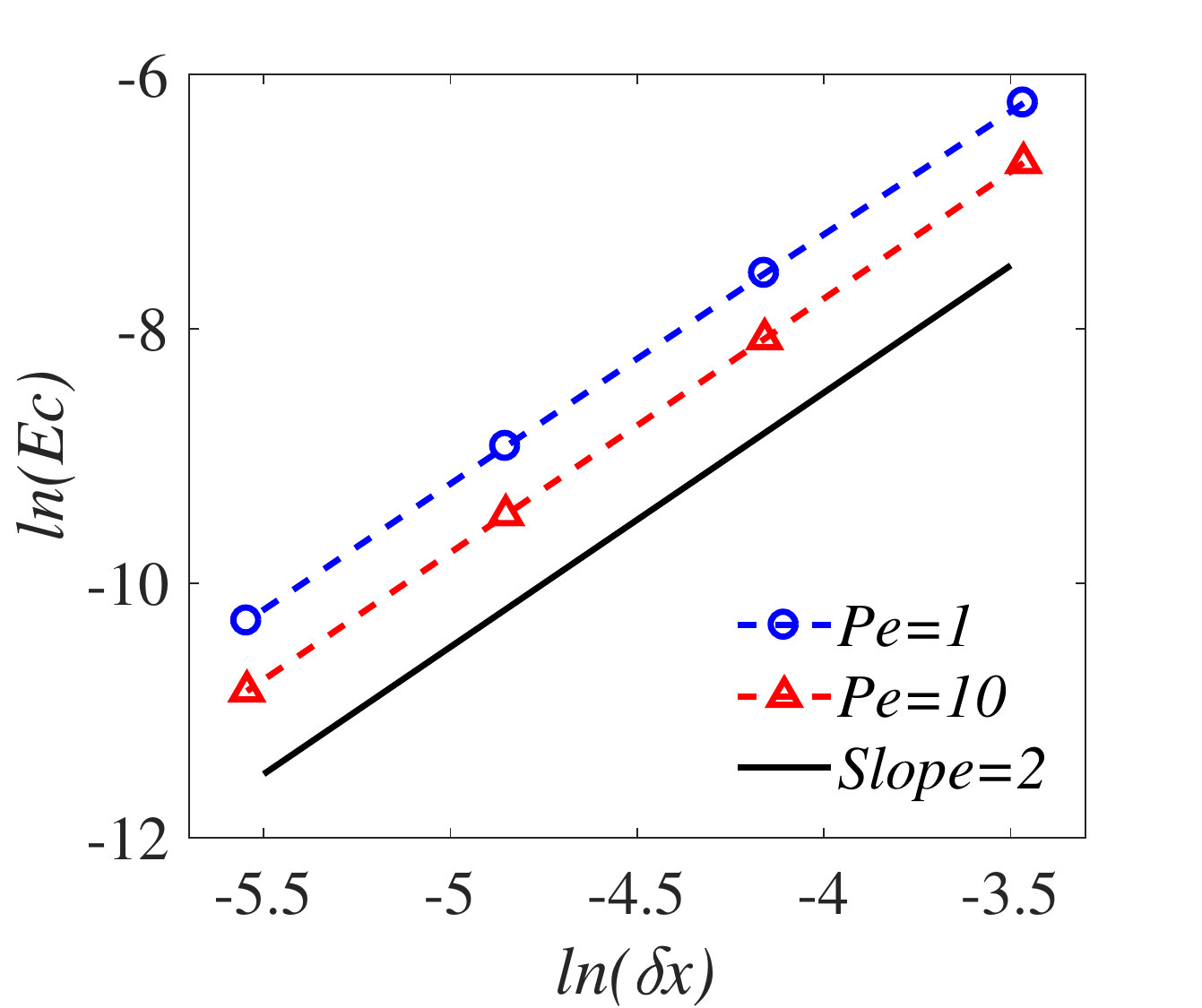}}~~
     \caption{The global relative errors of concentration ($E_c$) at different mesh sizes}
     \label{case1e2}
\end{figure}

The spatial accuracy of the proposed scheme is then tested, based on four grids with size of $32\times32$, $64\times64$, $128\times128$, $256\times256$. The global relative error of $C$ in the computational domain is defined as
 \begin{equation}
 E_c=\sqrt{\frac{\sum_{ij}(C-C^{*})^2}{\sum_{ij}(C^{*})^2}},
\label{eq:36}
\end{equation}
 where $C$ and $C^{*}$ are the concentration of the analytical solutions and numerical results respectively. The global relative error of $C$ in the computational domain at time $t = 1$ are plotted in~\cref{case1e2}, which clearly demonstrates the second-order accuracy of the scheme for straight walls.

\subsection{Convection-diffusion in an inclined channel}

\begin{figure}
 \centering
 \includegraphics[width=0.3\textwidth]{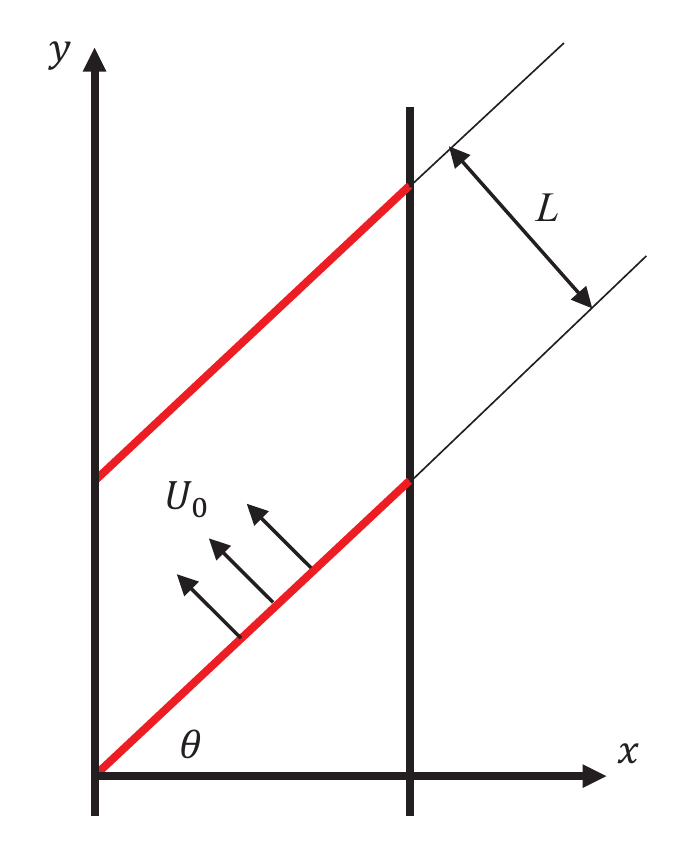}~
 \caption{Schematic of the inclined channel.}
 \label{case2e1}
\end{figure}

To demonstrate the applicability of the present scheme in complex geometries, a convection-diffusion problem in an inclined channel is studied. The configuration of the problem is shown in ~\cref{case2e1}. The fluid is injected into the domain with a constant vertical velocity $U_0$ through the bottom plate. $L$ is the distance between two parallel plates. $\theta$ is the inclination angle and a periodic domain is employed. A non-dimensional parameter is the P\'eclet number, which is define as $\text{Pe}=U_0L/D$. The concentration at the top plate are fixed at $C(x_{top},t) = 1.0$, while a Robin boundary condition is imposed on the bottom plate, (i.e., $\bm{n}\cdot \nabla C(x_{bottom},t)=a_1C(x_{bottom},t)+a_2$), the analytical solution for this case can be obtained in Ref~\cite{zhang2012general}

   \begin{equation}
  \begin{split}
   & C^{'}(x,y)=\frac{exp\{Pe(ycos\theta-xsin\theta)\}-a_1L+Pe}{exp(Pe)-a_1L+Pe},\\
   &u(x,y)=-U_0sin\theta,\\
   &v(x,y)=U_0cos\theta,
  \end{split}
\label{eq:37}
\end{equation}

\begin{figure}
     \centering
     \subfloat[]{\includegraphics[width=0.5\textwidth]{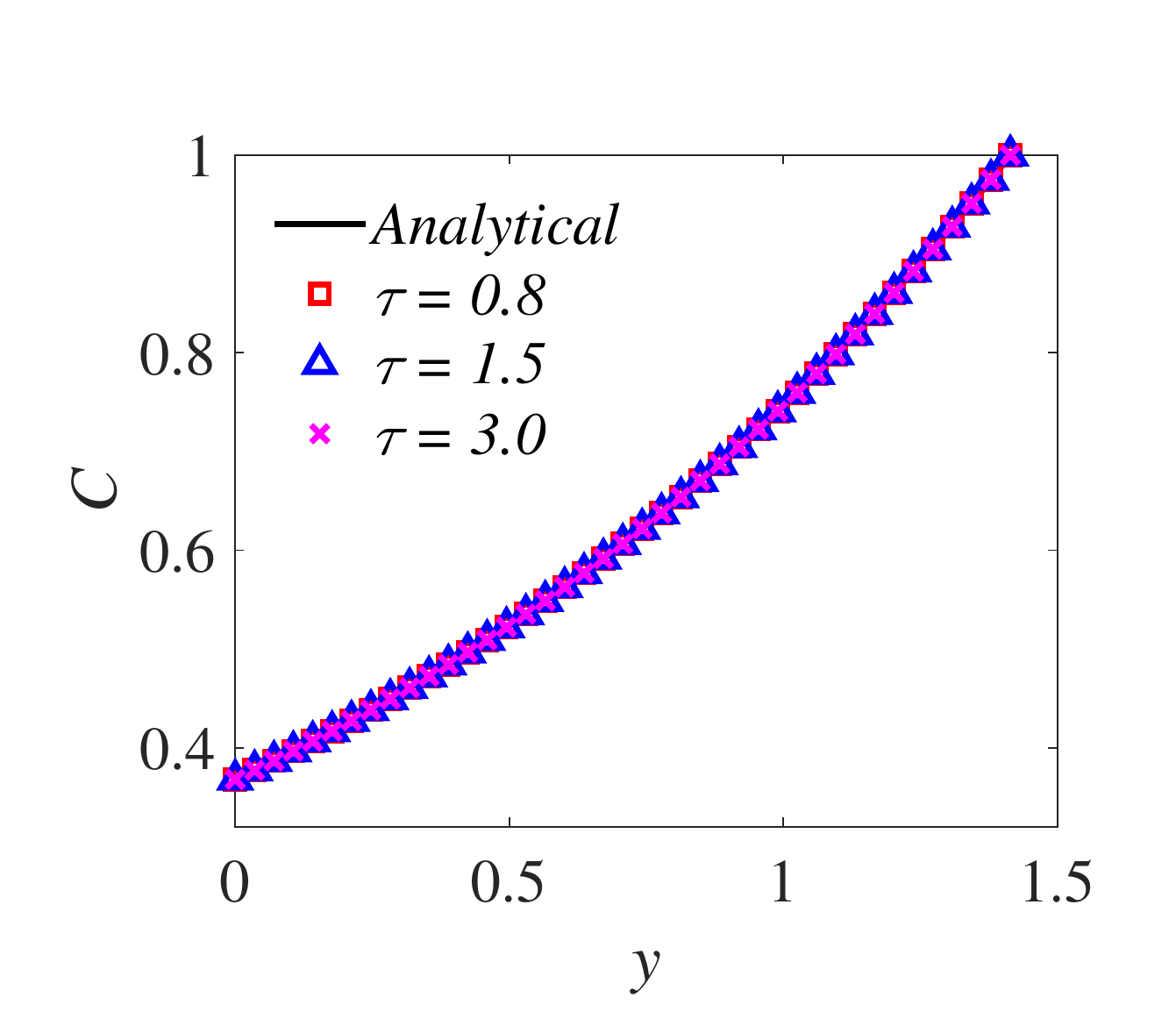}}~~
     \subfloat[]{\includegraphics[width=0.5\textwidth]{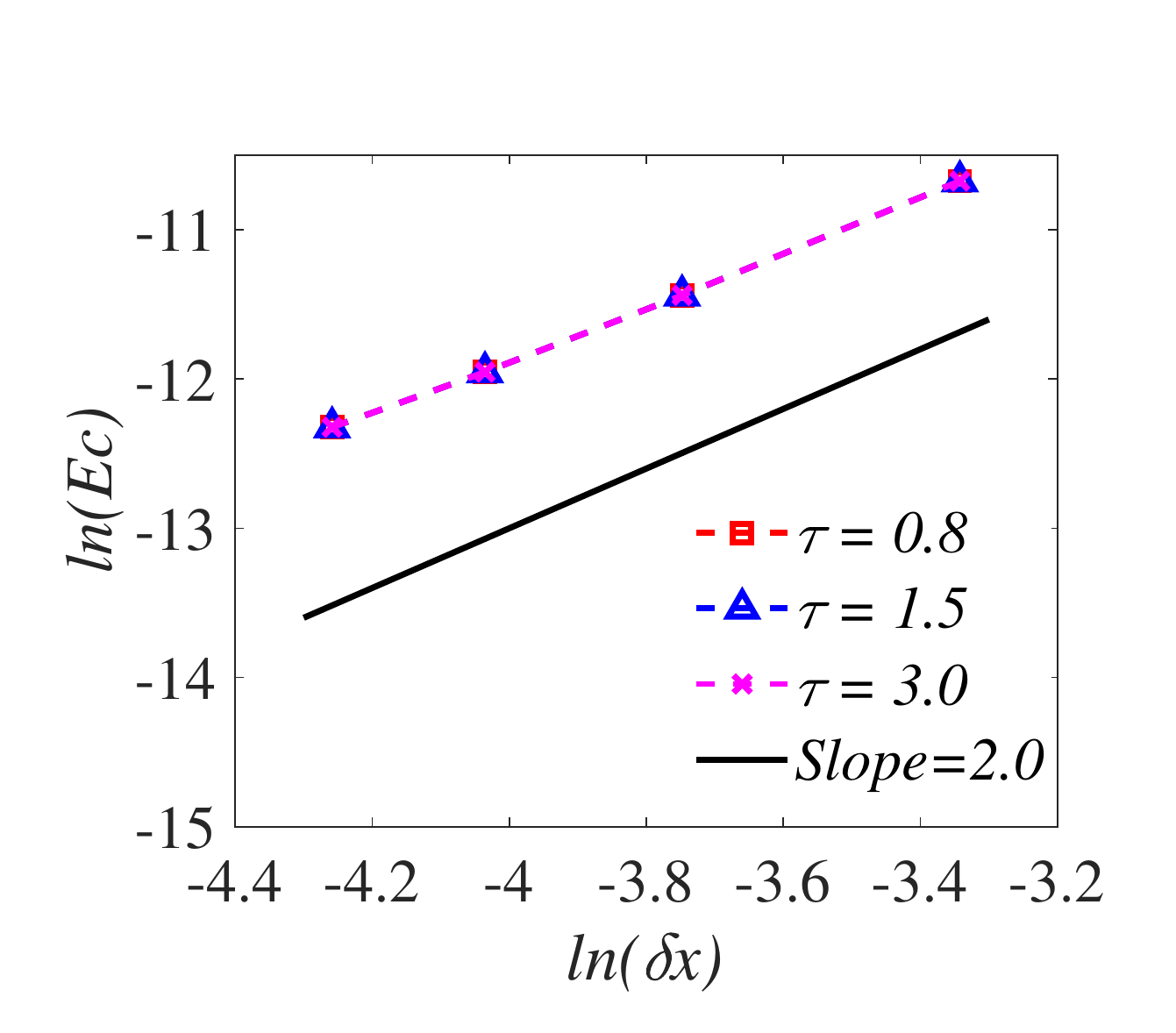}}~~
     \caption{Pe = 1, $\theta=\pi/4$. (a) concentration C along the line x = 0.5 through the inclined channel. (b) The global relative errors of concentration ($E_c$) at different mesh sizes }
     \label{case2e2}
\end{figure}

\subsubsection{\texorpdfstring{$\theta$}{[sigma]}=1/4}

We set $\theta=\pi/4$ to make all the lattice nodes locate on the boundary. The grid size is $40\times80$. Set $D=0.01$, $a_1=1,$ $a_2=0$ and the velocity as the same as the analytical one.

\begin{figure}
     \centering
     \subfloat[]{\includegraphics[width=0.5\textwidth]{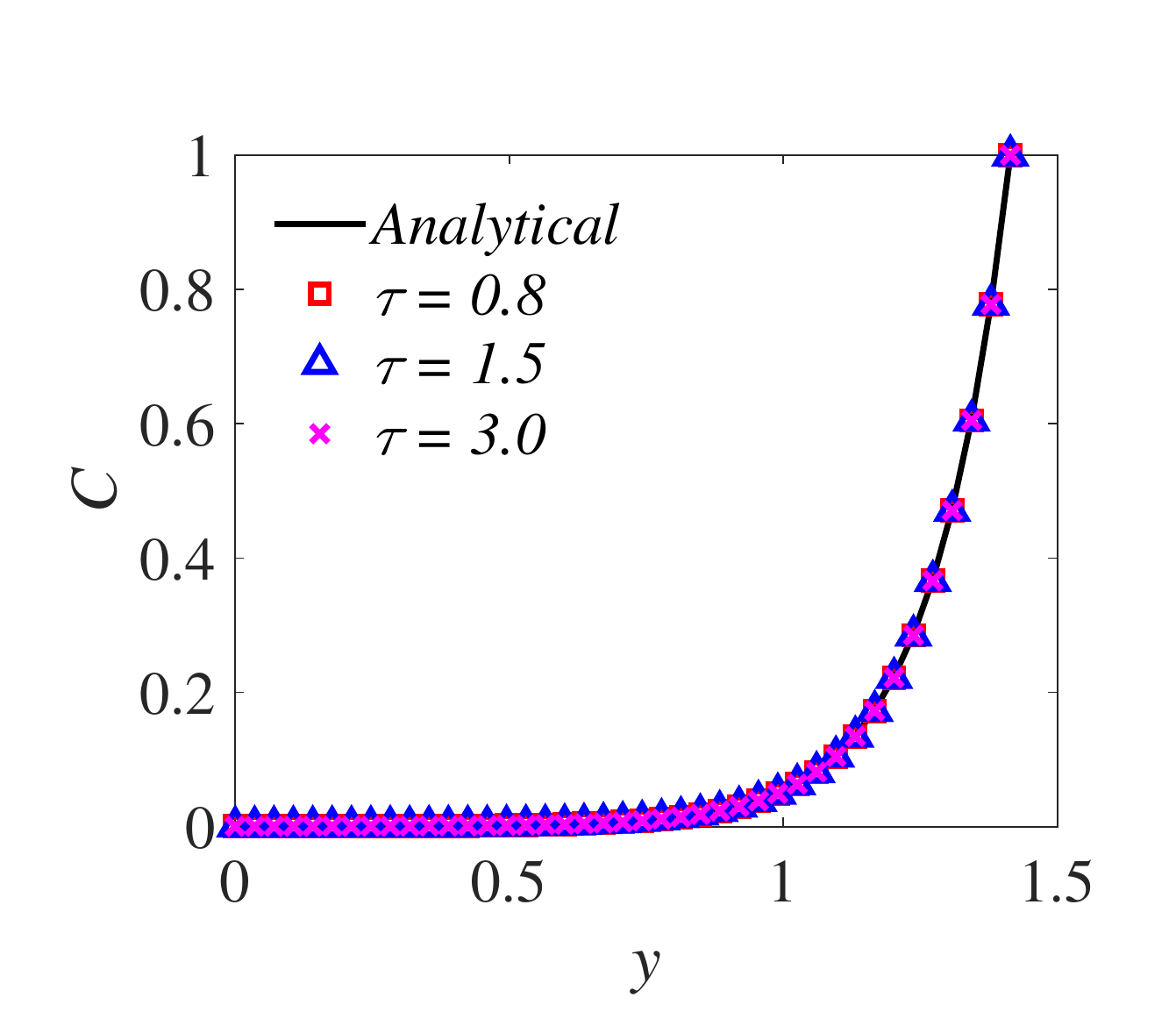}}~~
     \subfloat[]{\includegraphics[width=0.5\textwidth]{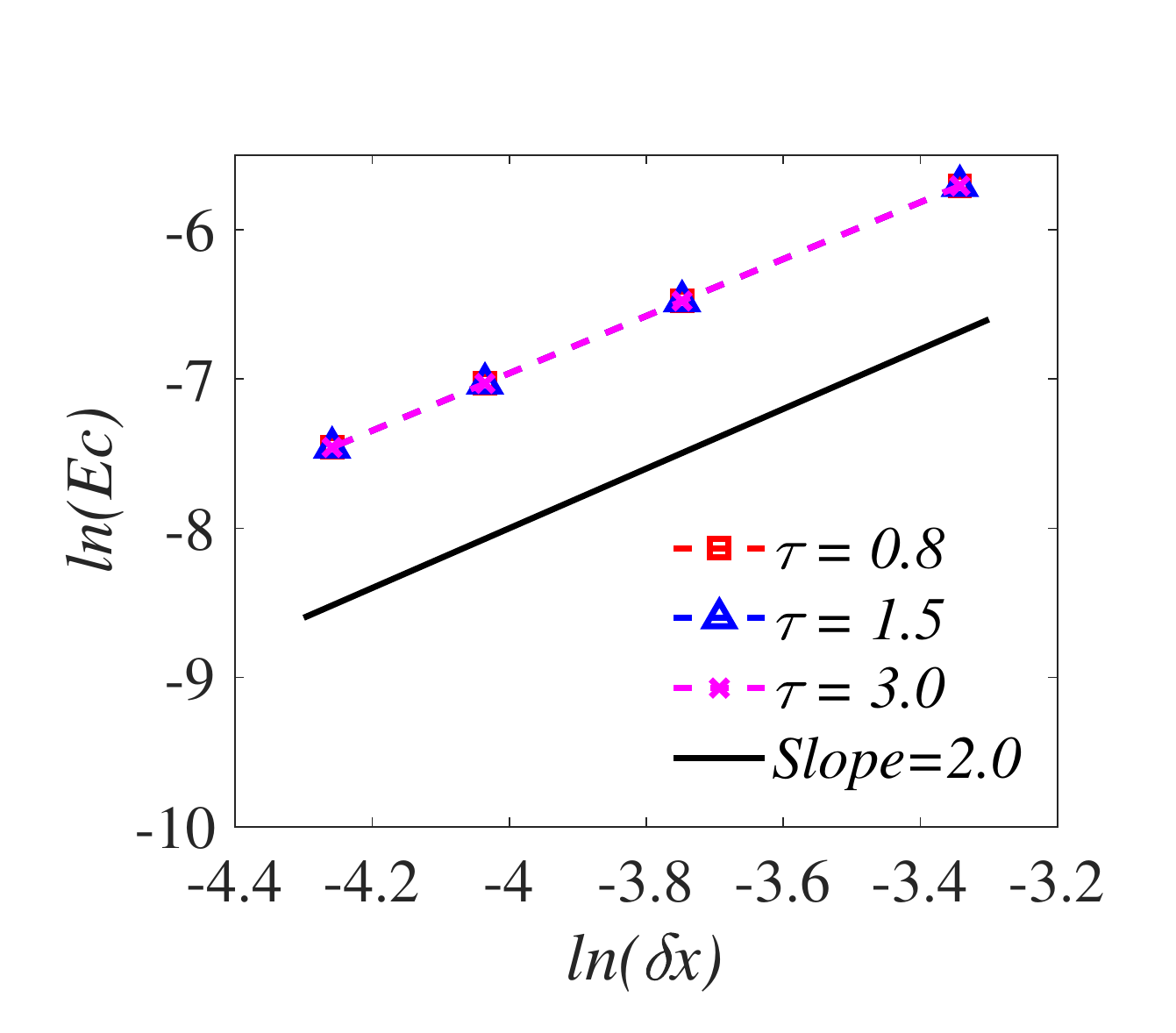}}~~
     \caption{Pe = 10, $\theta=\pi/4$. (a) Concentration C along the line x = 0.5 through the inclined channel. (b) The global relative errors of concentration ($E_c$) at different mesh sizes }
     \label{case2e3}
\end{figure}

We set $\text{Pe}=1$, and three different relaxation times (i.e., 0.8, 1.5, and 3.0) are used here. The profiles of concentration $C$ along the line $x=0.5$ through the inclined channel are shown in \cref{case2e2}a. It can be found that numerical results agree well with analytical solution. The relative global error of $C$ with different number of lattice spacing (i.e., $40\times80, 60\times120, 80\times160$ and $100\times200$) are shown in \cref{case2e2}b. It is clear that the present scheme has the second-order accuracy for this case.  The profiles and the relative global error of $C$ are also shown in \cref{case2e3} when $\text{Pe}=10$, as we can see that numerical results agree well with analytical solutions and it also has the second-order accuracy.

\subsubsection{\texorpdfstring{$\theta$}{[sigma]}=\texorpdfstring{$\arctan$}{[arctan]}(1/2)}

Furthermore, we set $\theta=arctan(1/2)$ to exam the accuracy of the scheme when the lattice nodes don't locate on the boundary. The grid size used in the simulations is $40\times40$ and other parameters are the same as when $\theta=\pi/4$.

\begin{figure}
     \centering
     \subfloat[]{\includegraphics[width=0.5\textwidth]{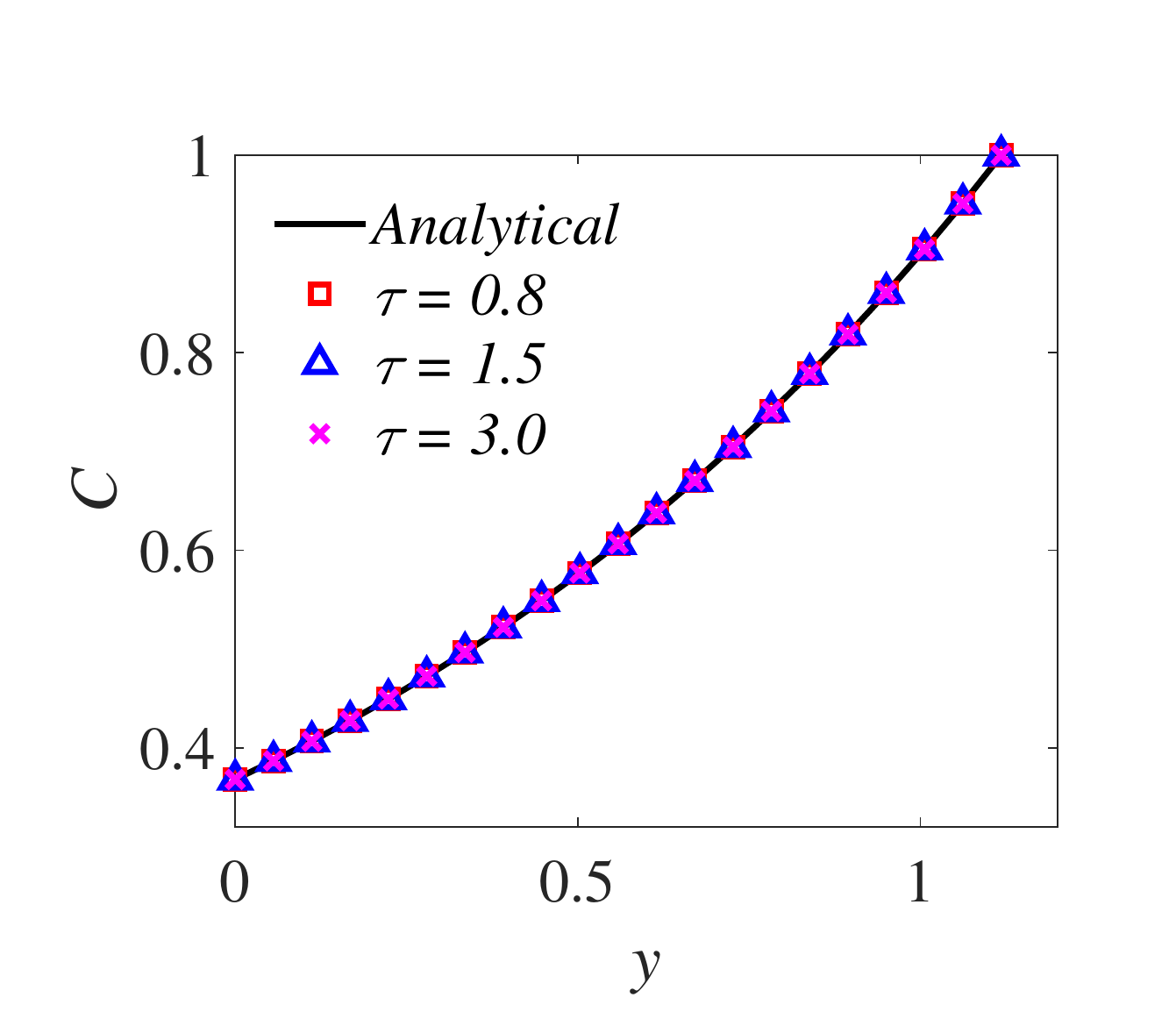}}~~
     \subfloat[]{\includegraphics[width=0.5\textwidth]{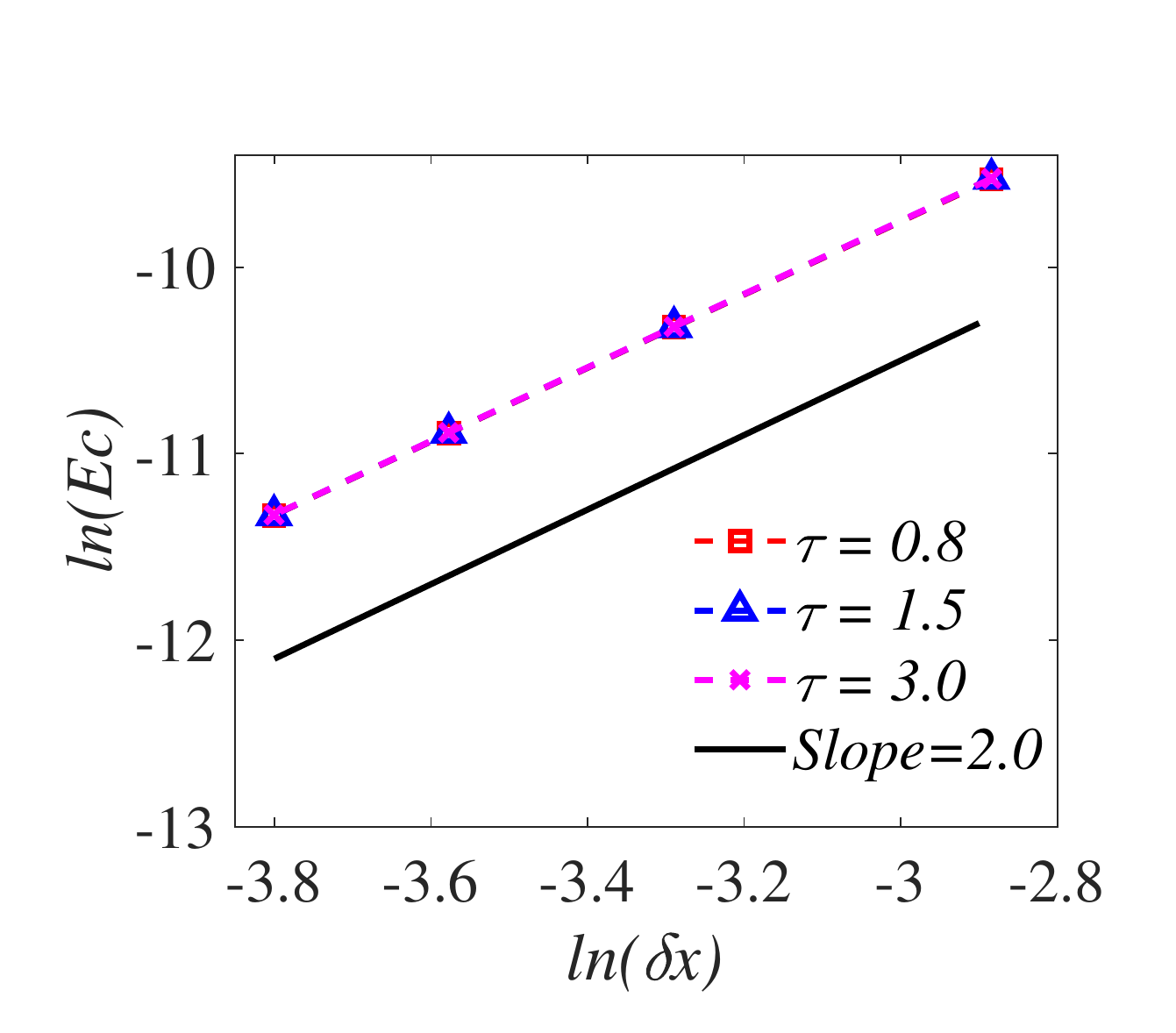}}~~
     \caption{Pe = 1, $\theta=arctan(1/2)$. (a) Concentration C along the line x = 0.5 through the inclined channel. (b) The global relative errors of concentration ($E_c$) at different mesh sizes }
     \label{case2e4}
\end{figure}

\begin{figure}
     \centering
     \subfloat[]{\includegraphics[width=0.5\textwidth]{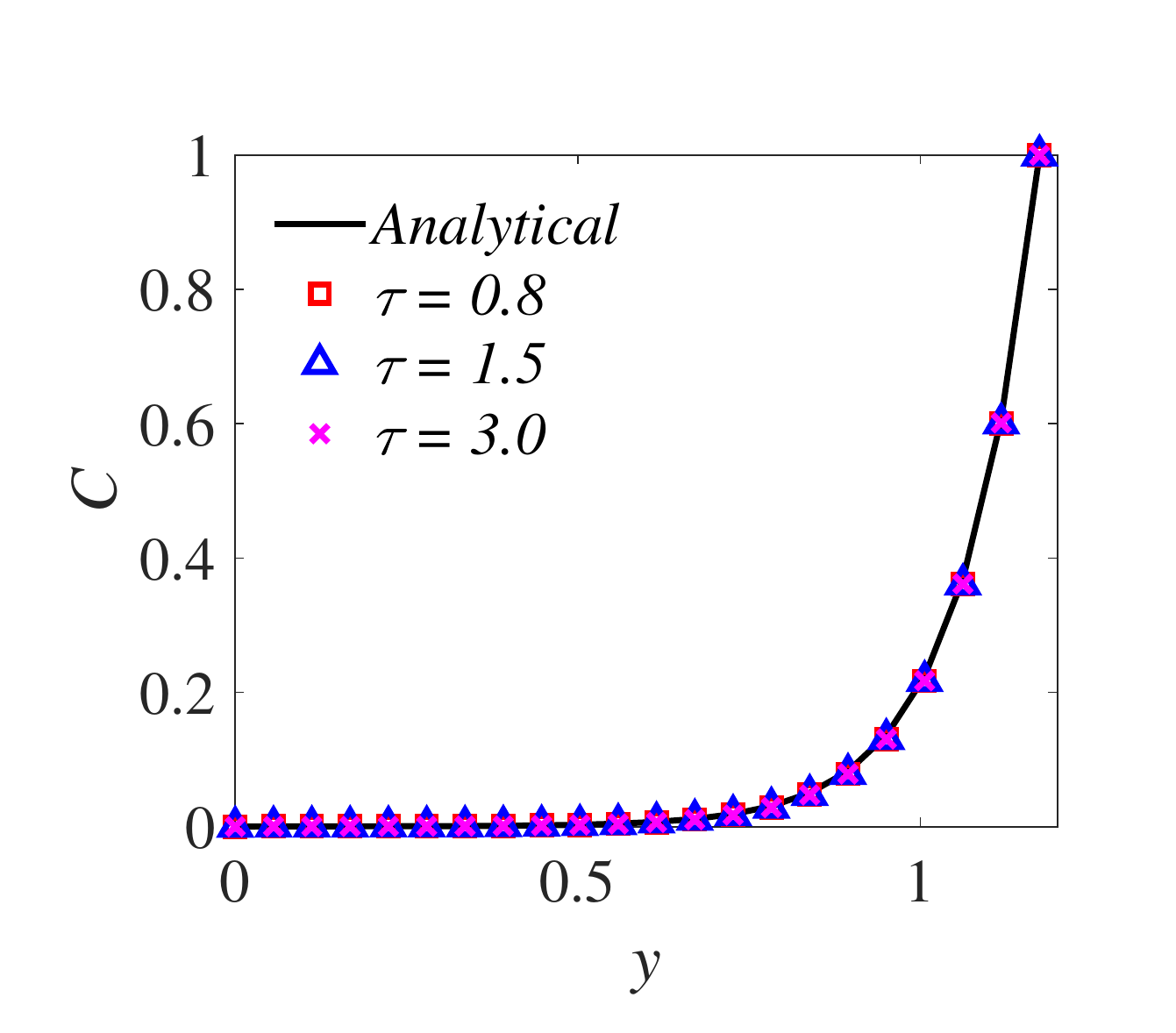}}~~
     \subfloat[]{\includegraphics[width=0.5\textwidth]{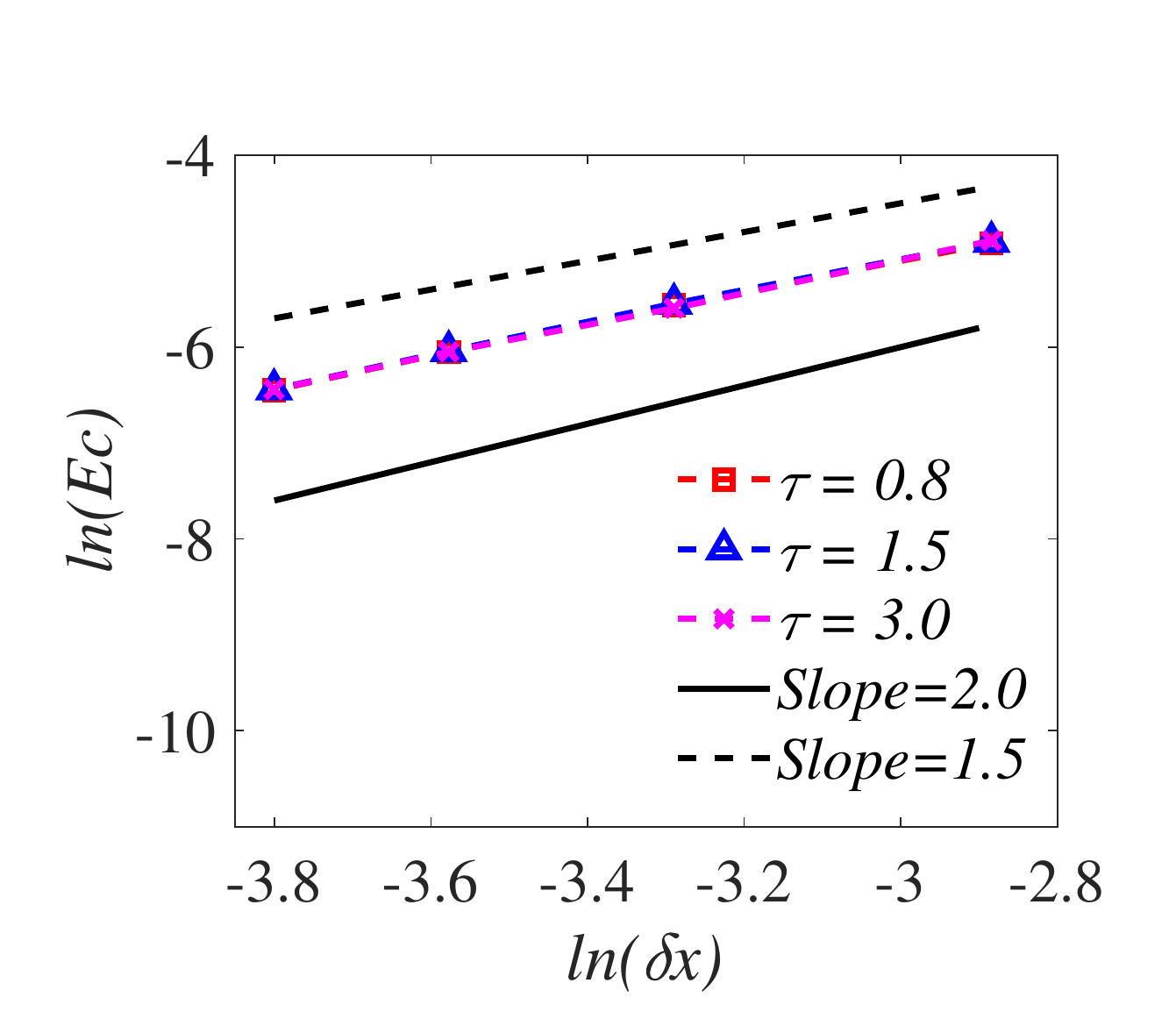}}~~
     \caption{Pe = 10, $\theta=arctan(1/2)$. (a) Concentration C along the line x = 0.5 through the inclined channel. (b) The global relative errors of concentration ($E_c$) at different mesh sizes }
     \label{case2e5}
\end{figure}

Set $\text{Pe}=1,10$. The profiles of concentration $C$ along the line $x=0.5$ through the inclined channel are shown in \cref{case2e4}a and \cref{case2e5}a. It is seen that the numerical results agree well with the analytical ones. The relative global error of $C$ with different number of lattice spacing (i.e., $40\times40, 60\times60, 80\times80$ and $100\times100$) are shown in \cref{case2e4}b and \cref{case2e5}b. When $\text{Pe}=1$, the present scheme has the second-order accuracy, but when $\text{Pe}=10$, the convergence order of the present is about 1.5. In other problems with complex geometries its order of accuracy maybe degenerated than this two case because the zig-zag boundaries approximation are used.
Although we can obtain their exact location by the methods of interpolation, the accuracy can be improved, while the implementation for curved boundaries are further complicated. So its useful to regard all the lattice nodes as the wall when dealing with complex boundary. The above results indicate that the proposed boundary scheme is capable of dealing with the linear heterogeneous surface reactions on complex boundaries.

\subsection{ Density driving flow with the dissolution reaction in cylindrical array}

\begin{figure}
 \centering
 \includegraphics[width=0.38\textwidth]{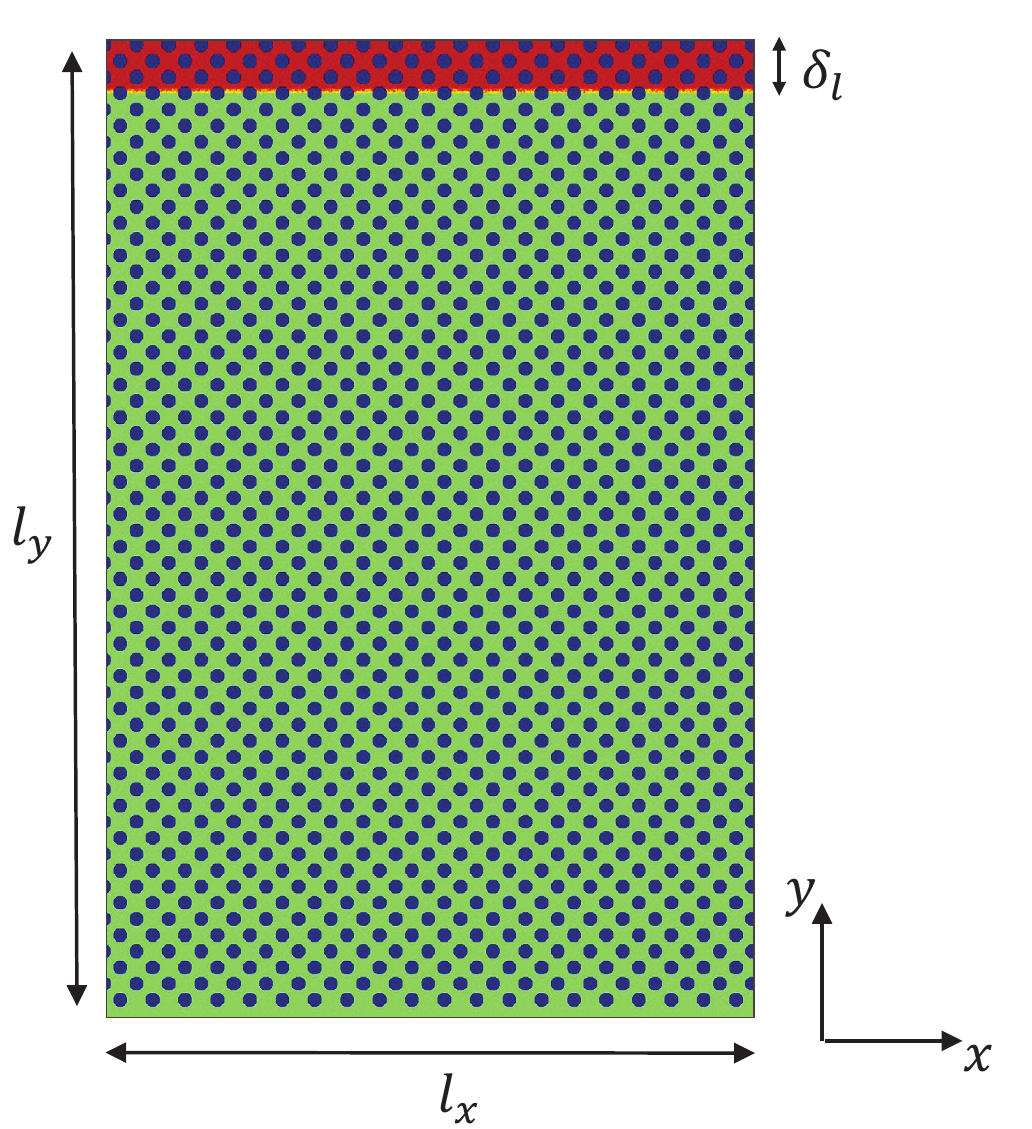}~
 \caption{Schematic of the cylindrical array.}
 \label{case3}
\end{figure}

In this section, density driving flow with the dissolution reaction in cylindrical array are simulated to demonstrate that our present scheme is applicable to complex practical problems.

Computational domain is displayed in \cref{case3}. The fluids whose concentration $C_1$ exist in the upper domain which occupies a length of $\delta_l$ with 100 meshes. The rest domain is occupied by the fluid whose concentration $C_2=0$. Then interface instability will occur because of the density difference and reactions between fluid and solid also occur at the same time. We set $l_x\times l_y=NX\times NY=1280\times 1936$, and radius of cylinders are $R=15$. We assume that the fluid is incompressible and it satisfies the Boussinesq approximation~\cite{spiegel1960boussinesq}. The governing equations of flow and solute transport can be expressed as
   \begin{equation}
  \begin{split}
   & \nabla\cdot\bm{ u}=0,\\
   &\partial_t\bm{u}+\bm{u}\cdot\nabla\bm{u}=-\frac{1}{\rho_0}\nabla p+\nu\nabla^2\bm{u}+\rho(C)\bm{g},\\
   &\partial_t\bm{C}+\bm{u}\cdot\nabla\bm{C}=D\nabla^2\bm{C},
  \end{split}
\label{eq:38}
\end{equation}
where $\rho(C)$ represents the density of the fluid which depends on its concentration: $\rho(C)=\rho_0[1+\eta(C-C_2)$. $\eta$ represent the density expansion coefficient which defined as $\eta=\partial\rho/\partial C$. The boundary conditions are given as
\begin{equation}
  \begin{split}
     & x=0, l_x:\quad \bm{u}=0,\quad\partial_nC=0,\\
     & y=0:\quad\bm{u}=0,\quad C=C_{in},\\
     & y=l_y:\quad\bm{u=0},\quad\partial_nC=0,
  \end{split}
\label{eq:39}
\end{equation}

If we take the characteristic parameters as:
\begin{equation}
  \begin{split}
     & L=l_x \qquad U=\sqrt{g\eta\Delta CL} \qquad  T=L/U \qquad C=C_1 ,
  \end{split}
\label{eq:40}
\end{equation}
Then this problem can be characterized by the Rayleigh number (Ra), Schmidt number (Sc) and Damk$\ddot{o}$hler number (Da):
\begin{equation}
  \begin{split}
     & \text{Ra}=\frac{g\eta L^3\Delta C_{max}}{\nu D},\quad \text{Sc}=\frac{\nu}{D},\quad \text{Da}=\frac{k_rL}{D}
  \end{split}
\label{eq:41}
\end{equation}
Previous studies~\cite{cardoso2014geochemistry,ward2014dissolution} have proved that the larger Ra is, the more likely it is to be instability. So in this section, we set Sc = 1, and $\text{Ra}=1\times10^7$ in order to observe the instability phenomenon clearly. Different value of Da are given to observe the effect of interface reaction on instability.
\begin{figure}
     \centering
     \subfloat[]{\includegraphics[width=0.2\textwidth]{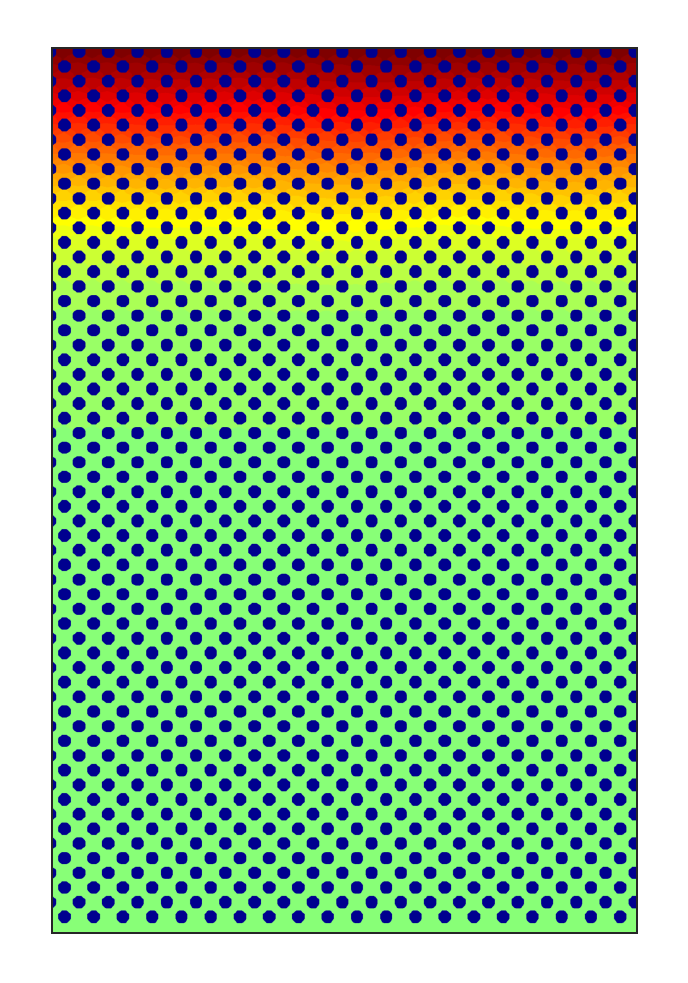}}~~
     \subfloat[]{\includegraphics[width=0.2\textwidth]{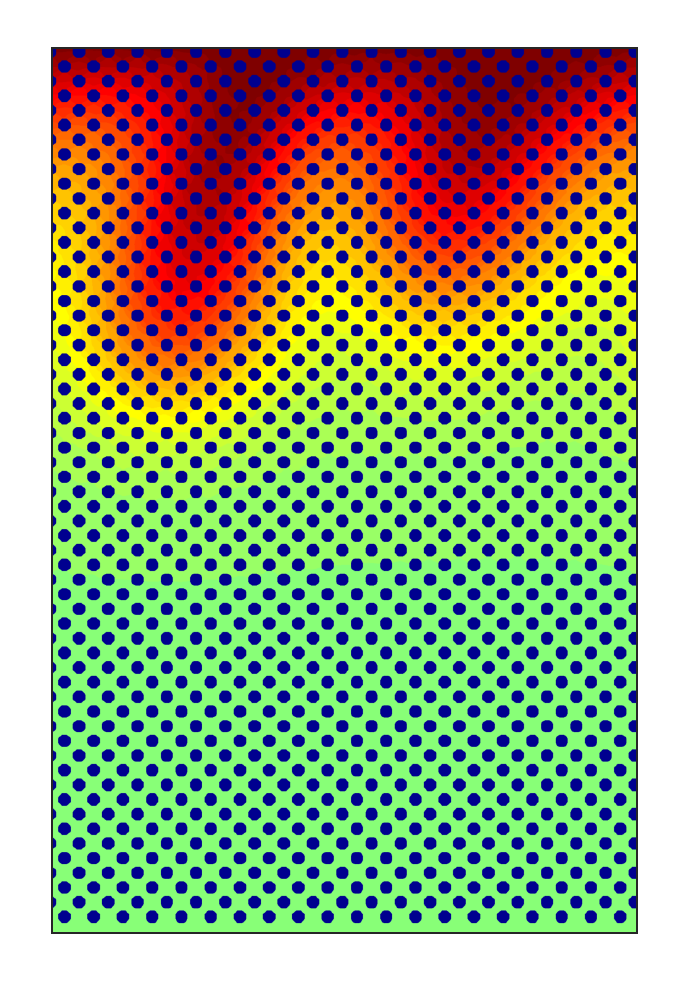}}~~
     \subfloat[]{\includegraphics[width=0.2\textwidth]{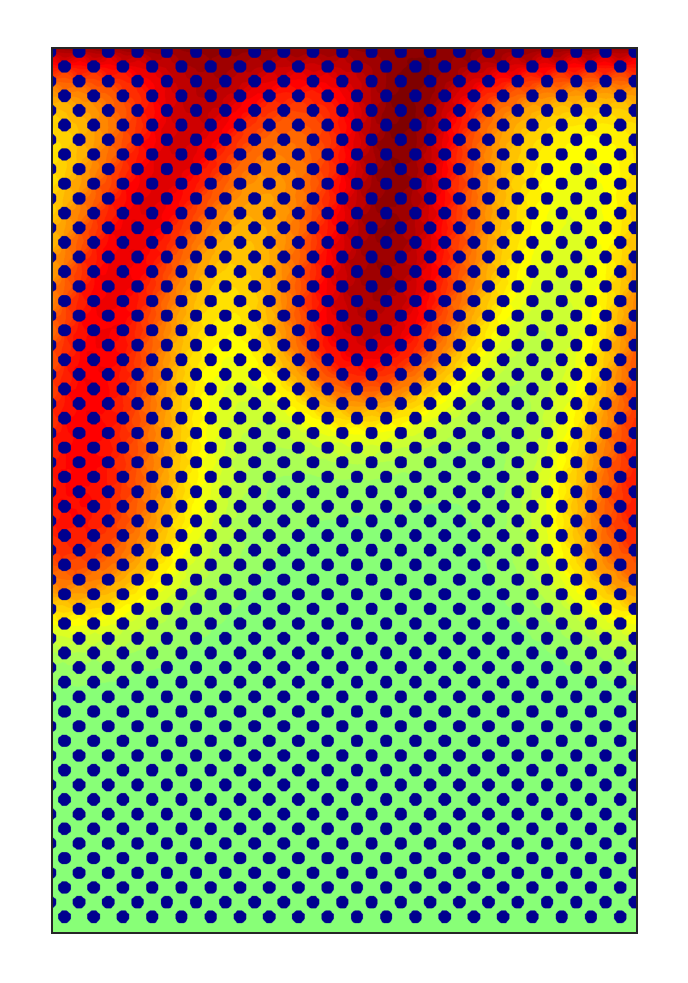}}~~
     \subfloat[]{\includegraphics[width=0.2\textwidth]{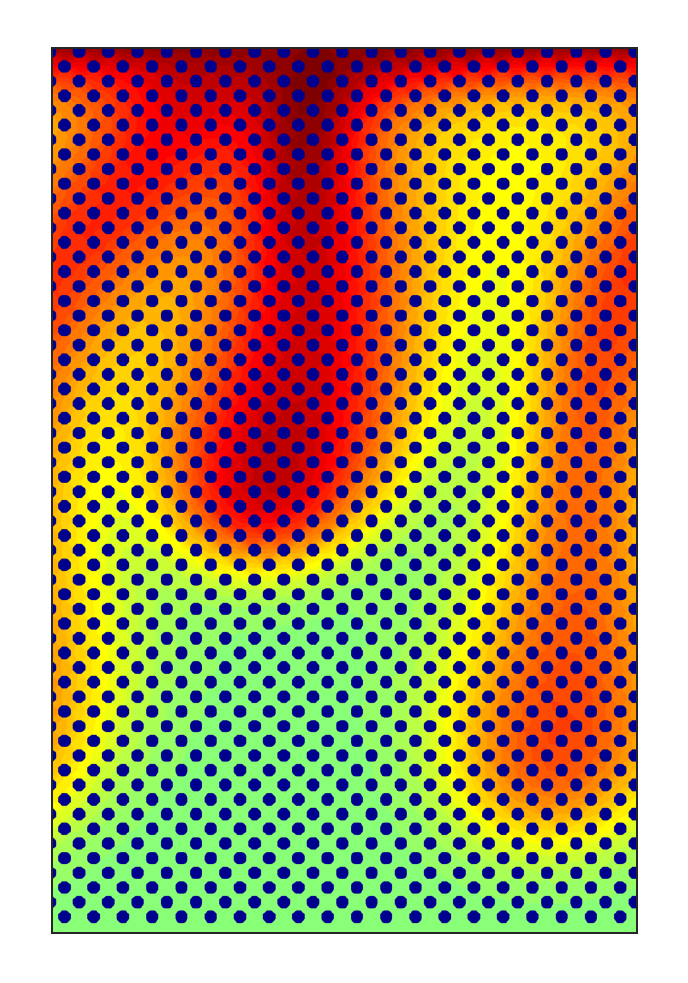}}~~
     \caption{Concentration profiles at different times when $\text{Da}=0$: (a) $t=200000$, (b) $t=350000$, (c) $t=450000$, (d) $t=550000$ }
     \label{case3e1}
\end{figure}

\begin{figure}
     \centering
     \subfloat[]{\includegraphics[width=0.2\textwidth]{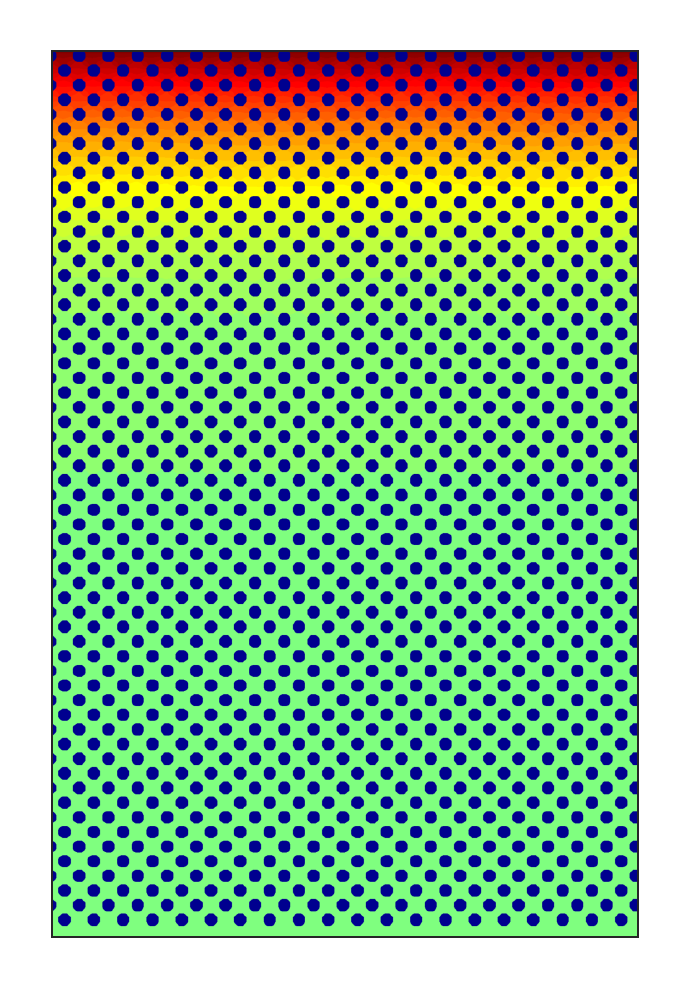}}~~
     \subfloat[]{\includegraphics[width=0.2\textwidth]{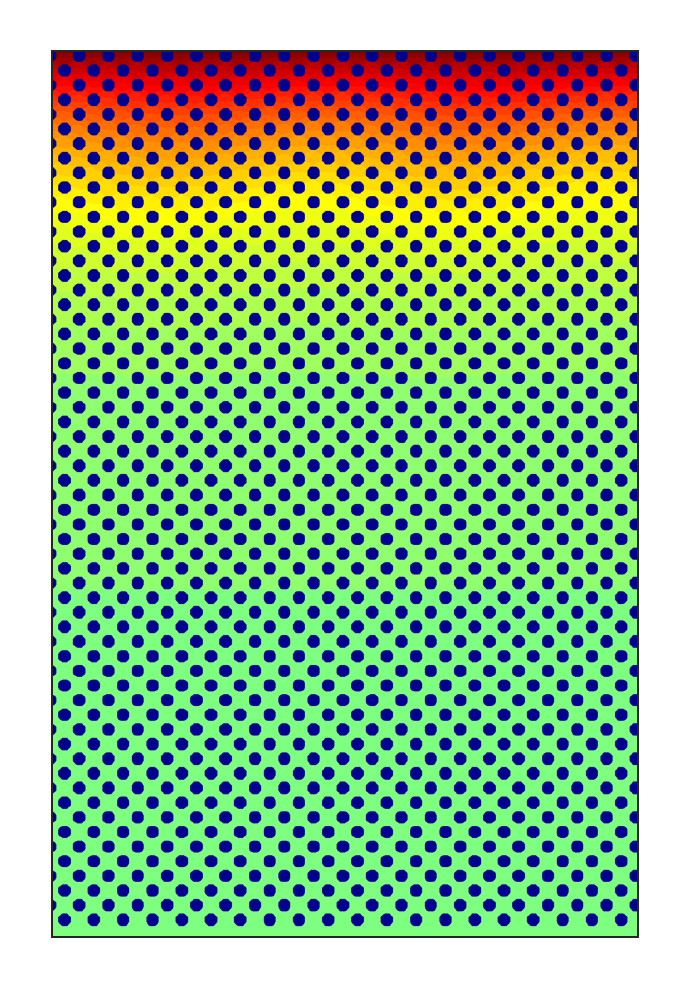}}~~
     \subfloat[]{\includegraphics[width=0.2\textwidth]{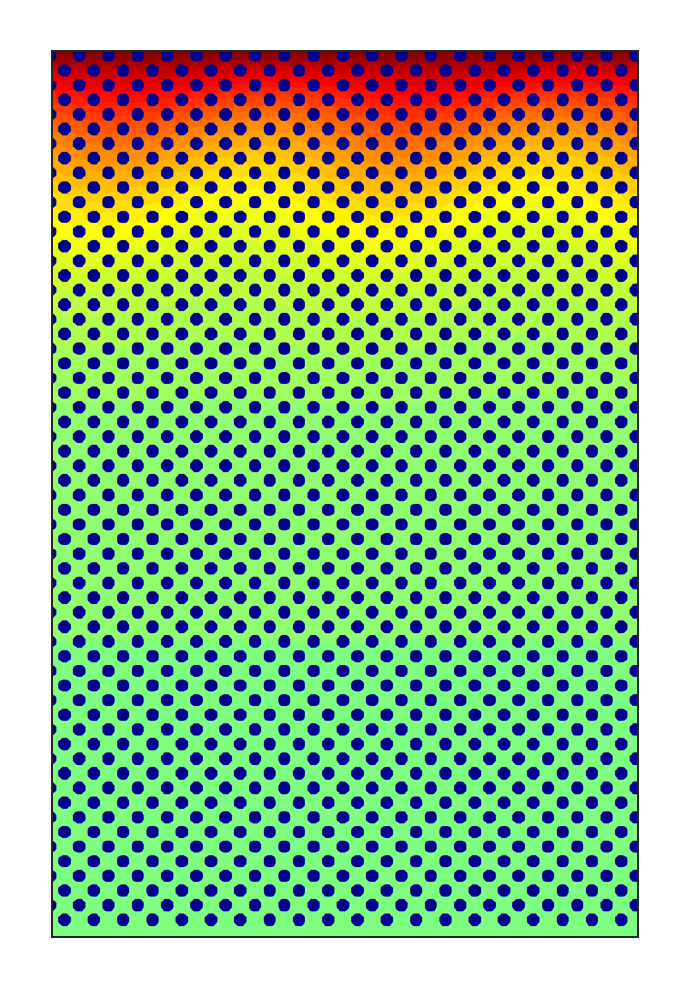}}~~
     \subfloat[]{\includegraphics[width=0.2\textwidth]{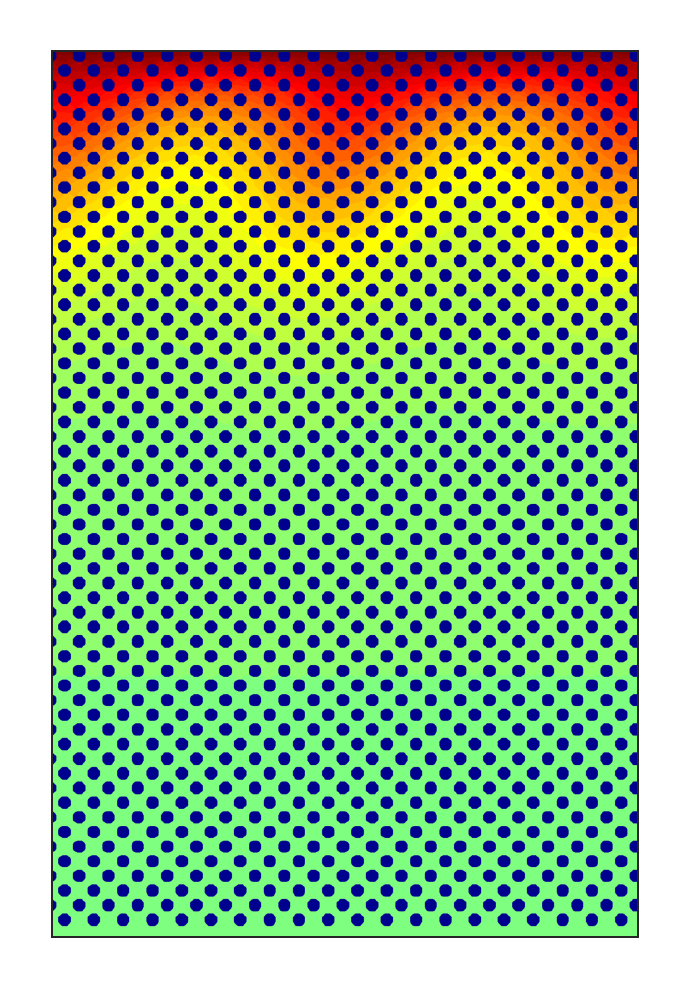}}~~
     \caption{Concentration profiles at different times when $\text{Da}=50$: (a) $t=200000$, (b) $t=350000$, (c) $t=450000$, (d) $t=550000$ }
     \label{case3e2}
\end{figure}

\begin{figure}
     \centering
     \subfloat[]{\includegraphics[width=0.2\textwidth]{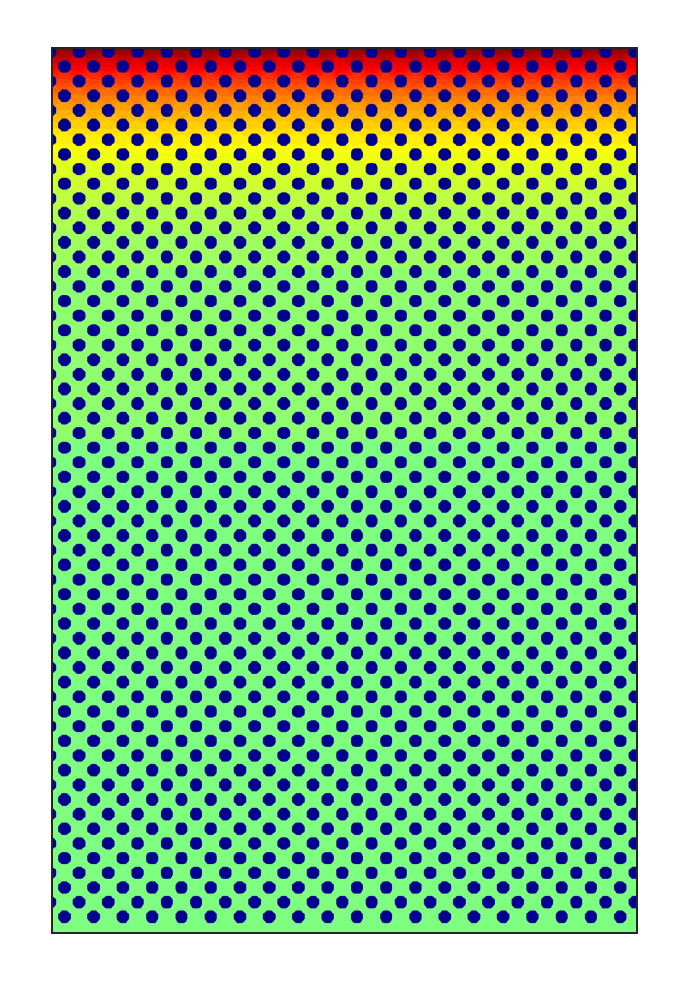}}~~
     \subfloat[]{\includegraphics[width=0.2\textwidth]{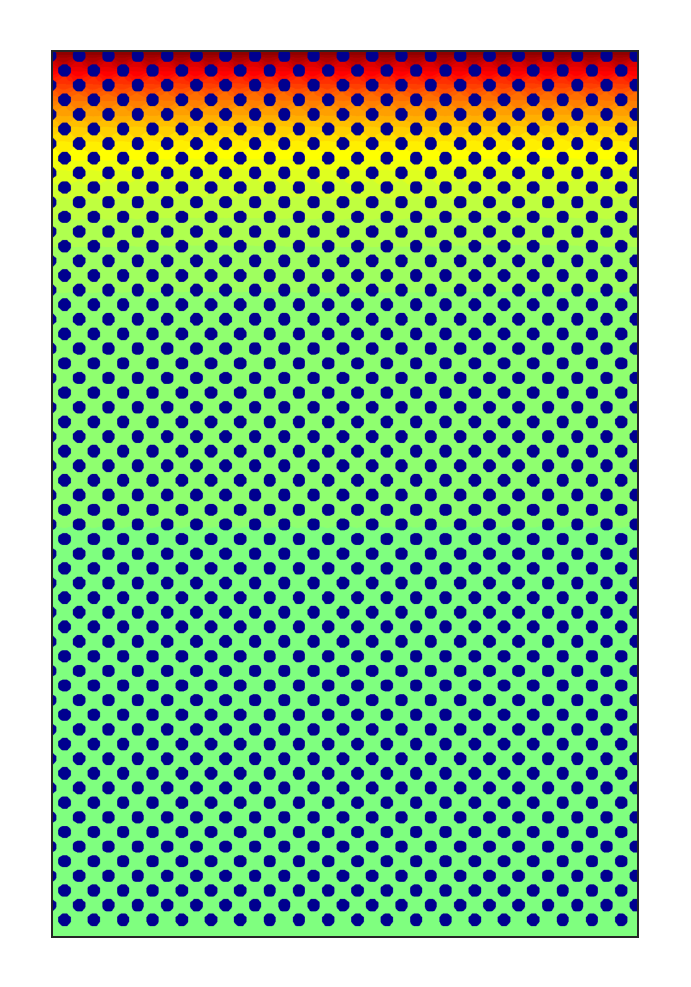}}~~
     \subfloat[]{\includegraphics[width=0.2\textwidth]{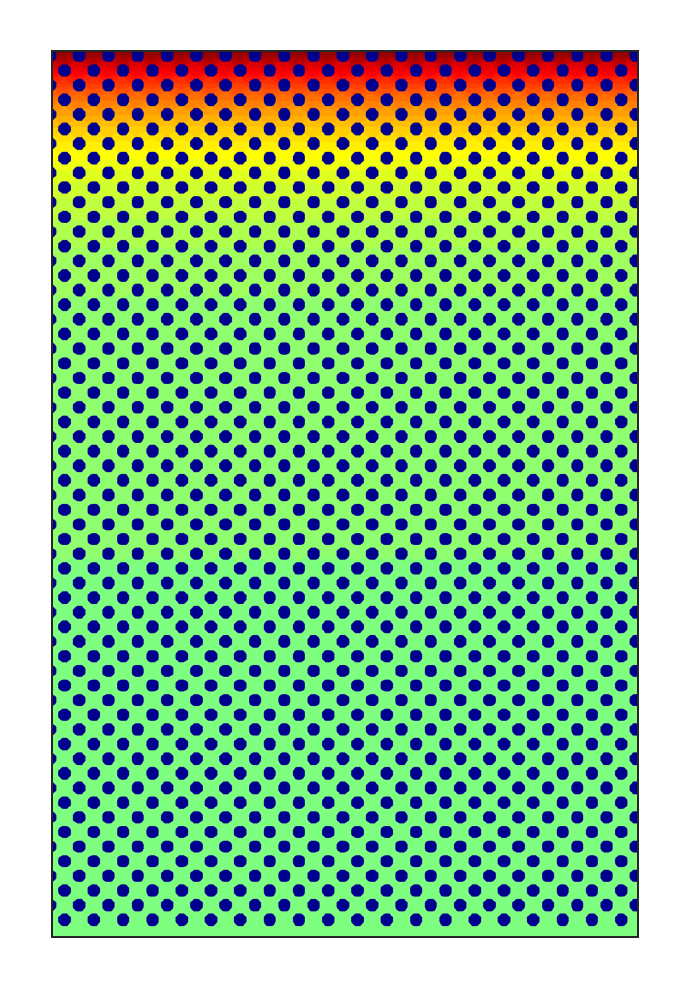}}~~
     \subfloat[]{\includegraphics[width=0.2\textwidth]{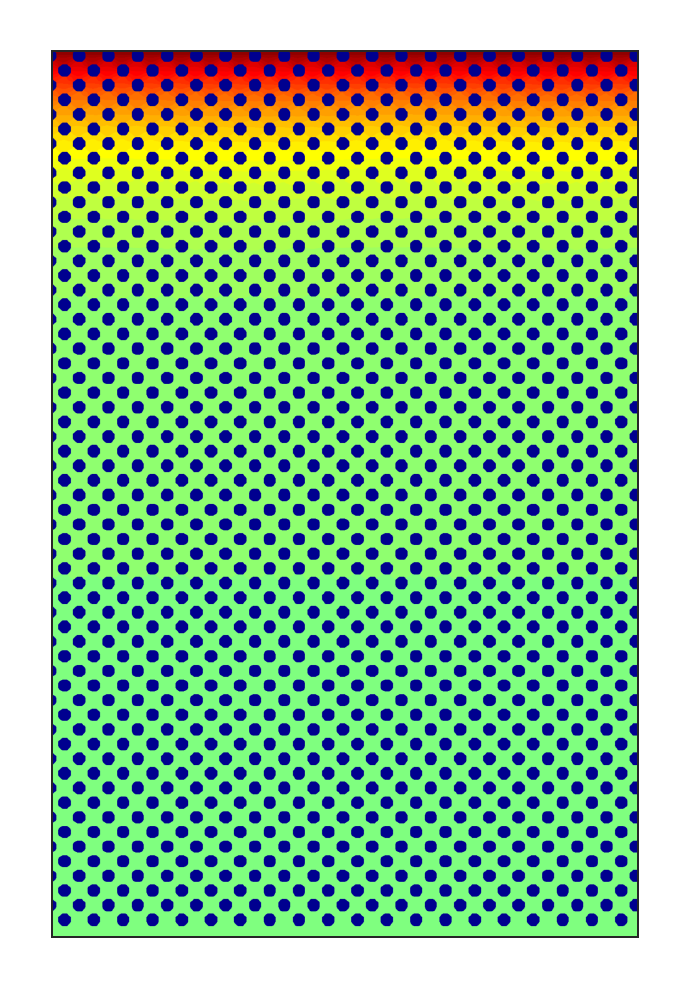}}~~
     \caption{Concentration profiles at different times when $\text{Da}=100$: (a) $t=200000$, (b) $t=350000$, (c) $t=450000$, (d) $t=550000$ }
     \label{case3e3}
\end{figure}

 First, we set Da=0, which means there is no reaction occur between the fluid and solid. \cref{case3e1} shows that when there is no chemical reaction, as time went by, the miscibility interface was no longer flat, and the heavy fluid was finger-like and penetrated into the light fluid because of the density difference. Then we set $\text{Da}=100, 1000$ to study the effect of chemical reactions on this process. As shown in \cref{case3e2} and \cref{case3e3}, the heterogenous reaction can suppress the instability, the heavy fluid stops moving down after it has dropped to a certain position over time. This is because when the chemical reaction rate is large enough, the amount of solvent consumed by the reaction reaches a dynamic equilibrium with the amount of solvent carried in the incoming flow. Therefore, the interface position does not change. So the denser fluid remain in the shallower regions of the formation for a long time because of the suppression of convection. It is same as the results of existing studies~\cite{cardoso2014geochemistry,ward2014dissolution} which demonstrates that our boundary scheme can solve practical complex problems.

\section{Conclusion}
In this paper, we propose a lattice Boltzmann boundary scheme for surface reactions. Compared to existing boundary schemes, our present scheme have a unified form for straight boundaries and curved boundaries. It consists of two parts, which represent the reaction of consumption and the reaction of formation respectively. So it has a clear physical implications. As the macroscopic equation on the boundary is avoided, the scheme can be easily implemented for complex geometry structures. In order to validate the accuracy of present model, we have performed numerical simulations for several problems, including convection-diffusion problems on both straight boundary and inclined boundary and a density driving flow with the dissolution reaction in cylindrical array. The results demonstrate that our model have second-order accuracy for straight boundaries. For curved boundaries its order of accuracy may be degenerated because the wall nodes mayn't locate on the boundaries.
In addition, our scheme can easily extended to three-dimensional LBE, it provide a more convenient tool to investigate reactive transport processes in natural and industrial applications

\section*{Acknowledgments}

\textcolor{red}{This work was supported by the National Key Research and Development Plan (No. 2016YFB0600805) and the National Science Foundation of China (11602091, 91530319).}

\section*{APPENDIX}

In this section, we discuss how the Eqs.~\eqref{eq:5} and ~\eqref{eq:16} can be obtained through the Chapman-Enskog expansion analysis~\cite{Meng2016boundary}. First, multiscale expansions can be introduced as follow:
\begin{align}
&g_i=g_i^{(0)}+\kappa g_i^{(1)}+\kappa^2g_i^{(2)}+\cdots, \tag{A1a} \\
&\partial_t=\kappa^2\partial_{t_2}, \tag{A1b}\\
&\nabla=\kappa\nabla_0. \tag{A1c},
\end{align}
where $\kappa$ here represents the expansion parameter. The Taylor's expansion is adopted to Eq.~\eqref{eq:2}, we can obtain that:
\begin{align}
D_ig_i+\frac{\delta_t}{2}D_{i}^{2}g_i=-\frac{1}{\tau\delta_t}\left[g_i-g_{i}^{eq}\right]. \tag{A2}
\end{align}
where $D_i=\partial_t+c_i\cdot\nabla.$ Then substituting Eqs.(A1) into Eq.(A2) and sort by the order of $\kappa$, we can obtain that:
\begin{align}
&\kappa^0: g_i^{(0)}=g_{i}^{eq}, \tag{A3a}\\
&\kappa^1: \bm{c}_i\cdot \bm{\nabla}_0 g_{i}^{(0)}=-\frac{1}{\tau\delta_t}g_{i}^{(1)},\tag{A3b}\\
&\kappa^2: \partial_{t_{2}}g_{i}^{(0)}+ \bm{c}_i\cdot \bm{\nabla}_0g_{i}^{(1)}+\frac{\delta_t}{2} (\bm{c}_i\cdot \bm{\nabla}_0)^{2}g_{i}^{(0)}=-\frac{1}{\tau\delta_t}g_{i}^{(2)}.\tag{A3c}
\end{align}
Substituting Eq.(A3b) into Eq.(A3c), we can get the following equation:
\begin{align}
 \partial_{t_{2}}g_{i}^{(0)}+ \bm{c}_i\cdot \bm{\nabla}_0\left[\left(1-\frac{1}{2\tau}\right)g_{i}^{(1)}\right]=-\frac{1}{\tau\delta_t}g_{i}^{(2)}.\tag{A4}
\end{align}
Then we take the zeroth lattice velocity moment of Eqs. (A3b) and (A4), we can obtain the following equations:
\begin{align}
& \bm{\nabla}_0\cdot (\bm{u}C)=0, \tag{A5a}\\
& \partial_{t_{2}}C-\bm{\nabla}_0\cdot\left[c_{s}^{2}(\tau-\frac{1}{2})\delta_t\bm{\nabla}_0C\right]=0,\tag{A5b}
\end{align}
Combining Eqs. (A5) and Eq. (A6), Eq.~\eqref{eq:1} can be recovered if $D$ is set as follow:
 \begin{align}
 D=c_s^2(\tau-0.5)\delta t,\tag{A6}
\end{align}
By taking the first lattice velocity moment of Eq. (A3b), the following equation can be obtained:
\begin{align}
 \bm{\nabla}_0C=-\frac{1}{c_{s}^{2}\tau\delta_t}\sum_{i}\bm{c}_ig_{i}^{(1)},\tag{A7}
\end{align}
from Eq. (A1a), we can obtain that $\kappa g_{i}^{(1)}\approx g_{i}^{neq}=g_i-g_{i}^{eq}$, so Eq.(A7) can be expressed as:
\begin{align}
 \bm{\nabla}C=-\frac{1}{c_{s}^{2}\tau\delta_t}\sum_{i}\left[\bm{c}_i(g_{i}-g_{i}^{eq})\right]=-\frac{1}{c_{s}^{2}\tau\delta_t}\left(\sum_{i}\bm{c}_ig_{i}-C\bm{u}\right),\tag{A8}
\end{align}
Combining Eq. (A6) and Eq. (A8), we can obtain:
\begin{align}
 \sum_{i}\bm{c}_ig_{i}-C\bm{u}=-c_{s}^{2}\tau\delta_t\frac{\partial C}{\partial \bm{n}},\tag{A9}
\end{align}
Comparing with Eq.~\eqref{eq:15}, then $\gamma$ in Eq.~\eqref{eq:15} can be obtained
\begin{align}
 \gamma=\frac{c_{s}^{2}\tau\delta_t}{D}=\frac{\tau}{\tau-0.5},\tag{A10}
\end{align}

\section*{References}
\bibliographystyle{IEEEtr}
\bibliography{ref}
\end{document}